\documentclass[a4paper,11pt]{article}
\pdfoutput=1 

\usepackage{jcappub} 

\usepackage[T1]{fontenc} 
\usepackage{graphicx,subfigure,multirow}
\usepackage{longtable}
\usepackage{dcolumn}
\usepackage{bm}
\usepackage[all]{xy}
\usepackage{color}
\usepackage{graphicx}
\usepackage[usenames,dvipsnames]{xcolor}
\usepackage{lscape}
\usepackage{tikz}
\usepackage{pgf}
\usepackage{booktabs}

\newcommand{\beq}{\begin{equation}}
\newcommand{\eeq}{\end{equation}}
\newcommand{\bea}{\begin{eqnarray}}
\newcommand{\eea}{\end{eqnarray}}
\mathchardef\minus="002D

\def\LCDM{$\Lambda \text{CDM }$}
\newcommand{\CLASS}{\texttt{CLASS}}

\title{Astrophysical and Cosmological Probes of Boosted Dark Matter}

\author[a]{Jeong Han Kim,}
\author[b]{Kyoungchul Kong,}
\author[a]{Se Hwan Lim,}
\author[c]{Jong-Chul Park}

\affiliation[a]{Department of Physics, Chungbuk National University, Cheongju, Chungbuk 28644, Republic of Korea}
\affiliation[b]{Department of Physics and Astronomy, University of Kansas, Lawrence, KS 66045, USA}
\affiliation[c]{Department of Physics and Institute of Quantum Systems, Chungnam National University, Daejeon 34134, Republic of Korea}

\emailAdd{jeonghan.kim@cbu.ac.kr}
\emailAdd{kckong@ku.edu}
\emailAdd{sehwan.lim@chungbuk.ac.kr}
\emailAdd{jcpark@cnu.ac.kr}

\abstract{
We present an in-depth study of two-component cold dark matter via extensive $N$-body simulations. 
We examine various cosmological observables including the temperature evolution, power spectrum, density perturbation, maximum circular velocity functions, and galactic density profiles. 
We find that a significant mass difference between the two components, coupled with the annihilation of the heavier into the lighter component, imparts warm dark matter (WDM)-like characteristics to the latter. 
This model benefits from the unique features of WDM, such as modifications to the matter power spectrum and density profiles, while avoiding stringent observational constraints on WDM mass. 
The two-component dark-matter model aligns with observational data and suggests new avenues for dark-matter detection in terrestrial experiments, particularly for light, sub-MeV DM candidates. 
Our findings provide a framework for understanding the small-scale structures and offer guidance for future particle physics and cosmological studies.
}

\begin{document}
\maketitle
\flushbottom

\section{Introduction}
\label{sec:intro}

Various astrophysical observations, such as galaxy rotation curves, galaxy clusters and gravitational lensing, have indisputably established the existence of dark matter (DM)~\cite{1980ApJ...238..471R, Corbelli:1999af, Refregier:2003ct, Allen:2011zs, Rubakov:2019lyf}. 
To further uncover the nature of DM, significant efforts have been made through direct detection experiments~\cite{Graham:2015ouw, Schumann:2019eaa}, indirect searches ~\cite{Gaskins:2016cha}, and investigations at the LHC~\cite{Boveia:2018yeb}. 
Yet, its properties remain largely enigmatic. 
Solving this long-standing puzzle in understanding DM could provide key insights into new physics beyond the standard model (SM)~\cite{Bertone:2016nfn}.

Cold dark matter (CDM), a key component of the cosmological \LCDM model, has successfully explained the large-scale structure of the Universe~\cite{Planck:2018vyg}. 
However, on smaller scales ($\lesssim 10~\text{kpc}$), CDM faces challenges~\cite{Perivolaropoulos:2021jda} such as the ``core-cusp problem'' \cite{deBlok:1997zlw, vandenBosch:1999ka, deBlok:2001hbg, KuziodeNaray:2007qi}, the ``missing satellites problem''~\cite{Klypin:1999uc,Moore:1999nt}, and the ``too-big-to-fail'' problem~\cite{Boylan-Kolchin:2011qkt, Garrison-Kimmel:2014vqa, Papastergis:2014aba}. 
While accurately accounting for baryonic feedback may alleviate some of these tensions to some extent~\cite{Wadepuhl:2010ib, Parry:2011iz, Governato:2012fa}, it remains uncertain if baryons are solely responsible~\cite{Buckley:2017ijx}.

These small-scale challenges, combined with null results from DM experiments, drive the exploration of new avenues beyond the CDM paradigm. 
One compelling hypothesis to address these issues is the possibility that dark sectors consist of more than one species. 
The simplest realization of this model is a mixture of CDM and warm dark matter (WDM), where the warm component can free-stream and dampen density perturbations on small scales. 
Well-motivated WDM candidates include gravitinos, sterile neutrinos, axinos, neutralinos, or any light thermally produced relic~\cite{Dodelson:1993je, Colombi:1995ze, ABAZAJIAN20171, ELLIS1984453, MOROI1993289, Gorbunov:2008ui, Shaposhnikov:2006xi, Hisano:2000dz}. 
Previous studies~\cite{Parimbelli:2021mtp, Anderhalden:2012jc, Maccio:2012rjx, Boyarsky:2008xj, Diamanti:2017xfo, Keeley:2023ive, Kamada:2016vsc} have extensively tested the mixed CDM-WDM model using various cosmological data and have derived constraints on the fraction of WDM as a function of its mass.

However, for free streaming to significantly impact astrophysical data, the mass of WDM must be on the order of $\mathcal{O}$(keV). 
To achieve similar effects for DM masses above the $\mathcal{O}$(keV) scale, an alternative mechanism is needed. 
One possibility is to introduce a mass hierarchy between two DM species and leverage their mass difference $\delta m$ to expel the lighter species through inter-conversion processes~\cite{Belanger:2011ww, Agashe:2014yua, DEramo:2010keq, Cheon:2008sym, Hambye:2008bq, Hambye:2009fg, Arina:2009uq, Belanger:2012vp, Carlson:1992fn, deLaix:1995vi, Hochberg:2014dra}. 
These lighter particles, given an extra boost - referred to as ``boosted DM'' - continue scattering with neighboring particles, thereby erasing matter perturbations on small scales.

In Refs.~\cite{Medvedev:2013vsa, Todoroki:2017pdh, Todoroki:2017kge, Todoroki:2020num}, the authors attempted the implementation of boosted DM at the macroscopic level using $N$-body simulations with a Monte Carlo approach~\cite{Burkert:2000di}.
They find that their two-component DM model with a highly degenerate mass spectrum $\delta m /m \sim 10^{-8}$ agrees with observational data at all scales due to substantial reduction of substructure and flattening of density profiles in the centers of DM halos found in simulations.

Despite its consistency with cosmological measurements, the followings remain unclear: how the phenomenology changes in a non-degenerate mass regime and how massive the DM particles can remain effective for this process. 
Generally, this issue intersects with the approach of treating DM particles as clumps in $N$-body simulations. 
This creates ambiguity in realizing the mass difference $\delta m$ between two DM species at the macroscopic clump level, as their direct translation is somewhat obscure.

Additionally, there has been limited discussion on the structure formation of boosted DM that considers the full history of cosmological evolution. 
The Monte Carlo approach adopted in Refs.
~\cite{Medvedev:2013vsa, Todoroki:2017pdh, Todoroki:2017kge, Todoroki:2020num} incorporates only the late-time effects of boosted DM at redshifts $z \lesssim 100$, excluding the early-time dynamics of perturbation evolution, which can significantly impact the late-time history. 
Ref.~\cite{Kamada:2021muh} examined the evolution of background cosmological quantities but used the Jeans wave number of light DM particles to roughly estimate the impact on structure formation. 
Although their rough estimations suggest a significant suppression of galactic abundance, the detailed evolution of density perturbations and the behavior of the matter power spectrum have not been computed.

Systematic research on the structure formation of boosted DM has only been conducted very recently, as highlighted in Ref.~\cite{Kim:2023onk}.
In this paper, we aim to provide more details and incorporate the following new points to gain a comprehensive understanding: 
\begin{enumerate}
    \item using linear perturbation theory, we systematically describe the growth of matter perturbations from the early Universe and compute the matter power spectrum today; 
    \item we incorporate nonlinear effects using $N$-body simulations, with initial conditions determined by linear perturbation theory predictions;
    \item we discuss various phenomenological aspects of the model beyond the degenerate mass regime; 
    \item we test the model's consistency with cosmological measurements and demonstrate its ability to address small-scale challenges; and 
    \item we derive constraints for the model using cosmological data and discuss future perspectives.
\end{enumerate}

The cosmological consequences of the boosted DM model are generically applicable and independent of its specific details. 
However, for a phenomenological study, we need to define a particular microscopic model, which will be discussed in Section~\ref{sec:twoDM}. 
In Section~\ref{sec:Evolution}, we presents a (semi-)analytical approach to the evolution of various cosmological quantities including background fields, temperature, perturbation, and the linear power spectrum. 
The impact on structure formation is examined in detail through full $N$-body simulations in Section~\ref{sec:nbody}, where we explore the nonlinear power spectrum, sub-halo structure, and galactic density profiles. 
In Section~\ref{sec:summary}, we provides a summary and discussion.

\section{Two-Component Dark-Matter Model}
\label{sec:twoDM}

We consider a simple model for a dark sector consisting of two Dirac fermions, $\chi_1$ and $\chi_2$, with a dark $U(1)'$ and $U(1)''$ gauge symmetries~\cite{Belanger:2011ww}. 
We assume that both $\chi_1$ and $\chi_2$ are charged under $U(1)''$, with a charge $+1$, while only $\chi_1$ is charged under $U(1)'$, with a charge $+1$. 
The dark sector is allowed to couple with the SM only through a kinetic mixing between $U(1)'$ and $U(1)_Y$.
These dark gauge symmetries are spontaneously broken by dark Higgses acquiring vacuum expectation values, which leads to massive dark photons $\gamma'$ and $\gamma''$ with masses $m_{\gamma'}$ and $m_{\gamma''}$ respectively. 
We are interested in the scenario where dark Higgses\footnote{To simplify the analysis, we will not take into consideration of mixing between the dark Higgs and the SM Higgs.} and dark photons are heavier than $\chi_1$ and $\chi_2$, with the mass hierarchy between $\chi_1$ and $\chi_2$, $m_2 > m_1$.
In this case, the heavier DM particle $\chi_2$ has no direct interaction with the SM, and annihilates only into the lighter DM particle $\chi_1$, while $\chi_1$ can directly annihilate into the SM particles.

As discussed in Section~\ref{sec:background}, heavy $\chi_2$ particles are kept to thermal equilibrium by the assist of light $\chi_1$ particles with a thermally-averaged cross section $\langle \sigma v \rangle_{22\rightarrow 11}$. 
Relic abundances of $\chi_1$ and $\chi_2$, denoted as $\Omega_{1}$ and $\Omega_{2}$ respectively, depend on the relative size of $\langle \sigma v \rangle_{22\rightarrow 11}$ and $\langle \sigma v \rangle_{11\rightarrow XX}$ where $X$ stands for SM particles.
The combined relic abundance of $\chi_1$ and $\chi_2$ should agree with the observed DM relic density, $\Omega_{\text{DM}} = \Omega_1 + \Omega_2 \simeq 0.27$. 
The fraction of $\chi_1$ is expressed by $r_1 = \Omega_1/ \Omega_{\text{DM}}$.

The free parameters of the model are
\begin{eqnarray} 
	\{ m_1, m_2, m_{\gamma'}, m_{\gamma''}, g', g'', \epsilon \} \;,
\label{eq:params1}
\end{eqnarray} 
where $g'$ and $g''$ denote gauge couplings for $U(1)'$ and $U(1)''$ groups respectively, and $\epsilon$ is the kinetic mixing parameter between $U(1)'$ and $U(1)_Y$.
Throughout this paper, we will trade the last four parameters in (\ref{eq:params1}), $\{m_{\gamma''}, g', g'', \epsilon \}$,
with $\{\Omega_{\text{DM}}, r_1, \sigma_{\text{self}1}, \sigma_{\text{self}2}\}$
where $\Omega_{\text{DM}}$ is fixed as $0.27$ to yield the observed DM relic density, and $\sigma_{\text{self}1}$ ($\sigma_{\text{self}2}$) denotes a self-scattering cross section of $\chi_1$ ($\chi_2$).  
We note that $\sigma_{\text{self}1}$ plays an important role in sharing the excessive kinetic energy of boosted $\chi_1$ particles with the rest of their species and hence increases the overall temperature. 
In principle, the self-interaction of non-relativistic $\chi_2$ particles can also influence on perturbation evolutions, but to focus on the main dynamics of $\chi_1$ self-heating, we will neglect the contribution of $\sigma_{\text{self}2}$ in our discussions.
Finally, the parameters which determine the most prominent phenomenology are
\begin{eqnarray} 
	\{ m_1, m_2, m_{\gamma'}, r_1, \sigma_{\text{self}1} \} \;.
\end{eqnarray} 
For concreteness, our discussions focus on the DM mass range of order 1 -- 100 MeV, 
$r_1$ = 0.1 -- 0.9, and $\sigma_{\text{self}1}$ = 0.1 -- 10 $\text{ cm}^2/\text{g}$, 
for a fixed value of $m_{\gamma'} = 30$ MeV.

\section{Analytical Insight on the Cosmological Evolution}
\label{sec:Evolution}

\subsection{Background Evolution}
\label{sec:background}

Cosmological evolution for the energy densities, $\rho_{1, 2, X}$, of $\chi_1$, $\chi_2$, and SM particles $X$ is governed by the following coupled Boltzmann equations,
\begin{eqnarray} \label{eq:BE01}
    \frac{d \rho_{2}}{dt} + 3H \rho_{2} =&-&\frac{\big<\sigma v \big>_{22\rightarrow 11}}{m_{2}} \left( \rho^2_{2} 
    - \frac{\rho^2_{2,\text{eq}}}{\rho^2_{1,\text{eq}}} \rho^2_{1} \right) \;,\\ 
\label{eq:BE02}
    \frac{d \rho_{1}}{dt} + 3H \rho_{1} = &-&\frac{\big<\sigma v \big>_{11\rightarrow X X}}{m_{1}} \Big( \rho^2_{1} 
    -\rho^2_{{1,\text{eq}}} \Big)
    + \frac{m_{1}}{m_{2}} \frac{\big<\sigma v \big>_{22\rightarrow 11}}{m_{2}} \left( \rho^2_{2} 
    - \frac{\rho^2_{2,\text{eq}}}{\rho^2_{1,\text{eq}}} \rho^2_{1} \right) \;, \\ 
  \label{eq:BE03}
 \frac{d \rho_{X}}{dt} + 4H \rho_{X} = && \frac{\big<\sigma v \big>_{11\rightarrow X X}}{m_{1}} \Big( \rho^2_{1} - \rho^2_{{1,\text{eq}}} \Big)  \;.
\end{eqnarray}
Here the Hubble parameter $H$ is determined by the Friedmann equation $H^2 = 8 \pi G \rho_{\rm{tot}} / 3$, where $\rho_{\rm{tot}}$ denotes the total energy density of the Universe and $G$ is Newton's gravitational constant%
\footnote{When relativistic species dominate the Universe during early times, the Friedmann equation reads $H^2 \simeq g_* \pi^2 T^4/(90 M^2_{\rm{pl}})$ where $g_*$ denotes the effective number of degrees of freedom at the temperature $T$ and $M_{\text{pl}} = 1/\sqrt{8\pi G}$ is the reduced Planck mass.}.
The energy density in thermal equilibrium is given by ($M = 1, 2, X$)
\begin{eqnarray}  
\rho_{M, {\text{eq}}} &=& \int \frac{d^3p}{(2\pi)^3} E_{M} ~e^{-E_{M}/T} \\
&=& \frac{m^4_{M}}{2 \pi^2} \left( \frac{T}{m_{M}} \right) \left[ K_1 (m_{M}/T) + 3\left( \frac{T}{m_{M}}\right) K_2 (m_{M}/T) \right] \;, \nonumber
\end{eqnarray} 
where $K_{n}$ are the modified Bessel functions of the second kind and $T$ is the temperature of the radiation bath. 
For relativistic particles, the replacements of $m_{M} \rightarrow \langle E_{M} \rangle$ are required in Eqs.~(\ref{eq:BE01}--\ref{eq:BE03}), where $\langle E_{M} \rangle= \rho_{M}/n_{M}$ refers to the average energies of dark matter particles, while $n_M$ represents their number density.

\begin{figure}[t]
    \centering
    \includegraphics[width=0.49\linewidth,clip]{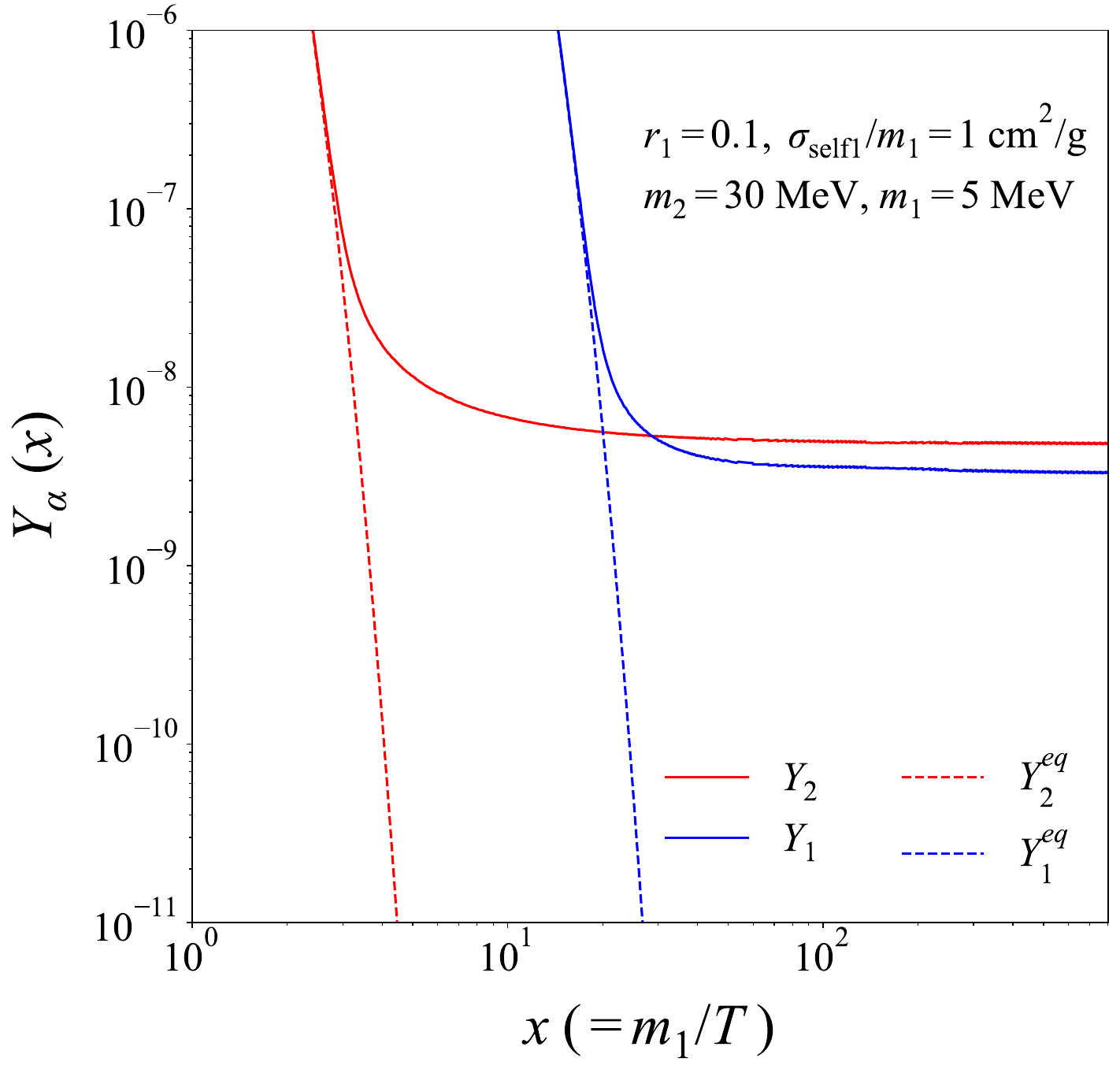}
    \includegraphics[width=0.49\linewidth,clip]{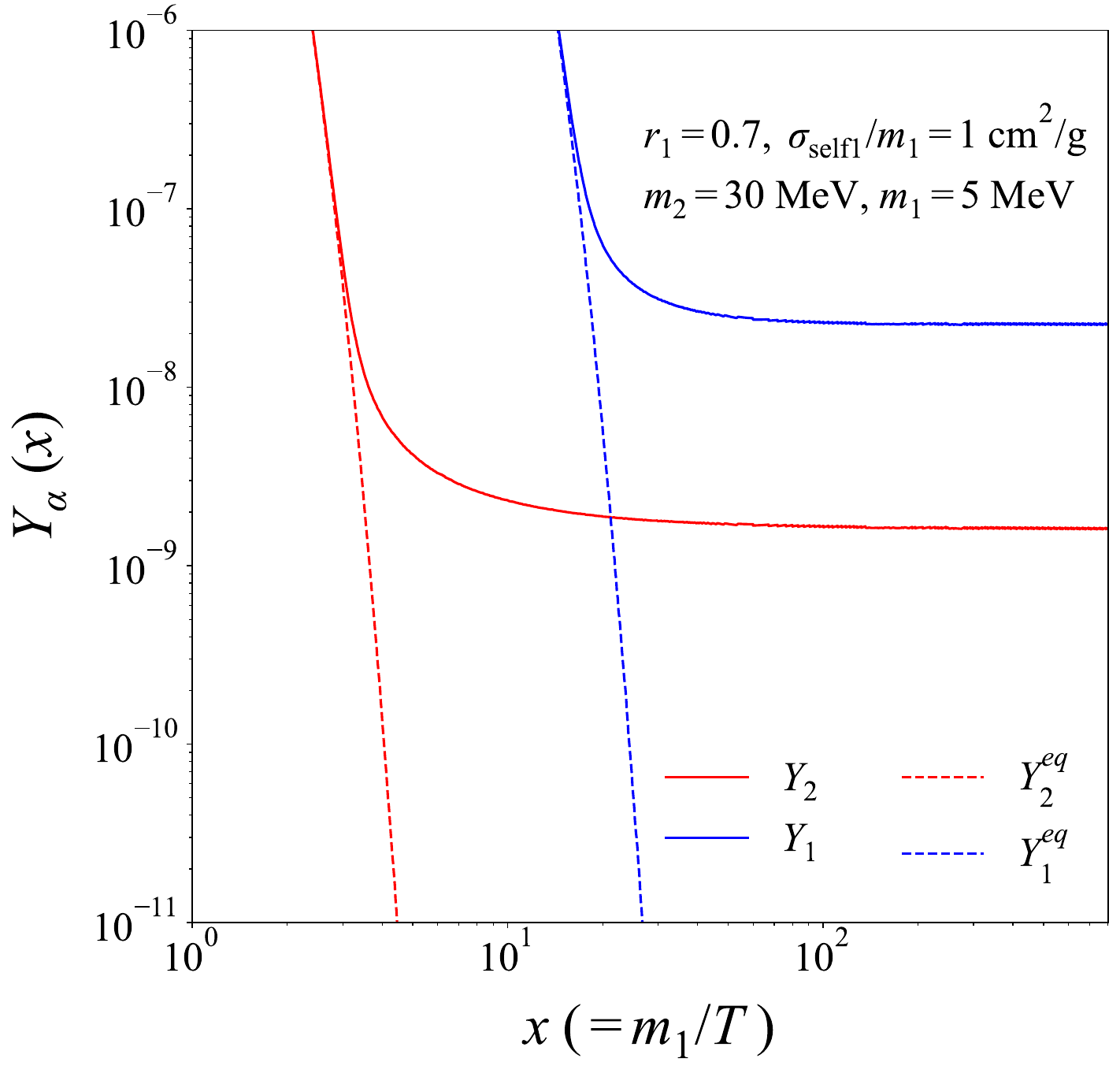}
    \caption{The evolutions of comoving number densities (solid lines), $Y_1$ (blue) and $Y_2$ (red), for $r_1 = 0.1$ (left) and $r_1 = 0.7$ (right) with fixing $m_1 = 5$ MeV and $m_2 = 30$ MeV. Dashed lines denote the  corresponding equilibrium abundances.  Here, we have introduced the ratio parameter $x=m_{1}/T$ where $T$ denotes the temperature of the radiation bath.
    }
    \label{fig:Relics}
\end{figure}

To solve the coupled Boltzmann equations (\ref{eq:BE01}--\ref{eq:BE02}), it is customary to introduce the variable $Y=n/s$, which are related to the numbers of particles in a comoving volume $N_M = n_M /s \sim Y_M$, where $s=2\pi^2 g_\ast T^3/45$ is the total entropy density. 
We also show our numerical results in terms of $Y_M$ in Figure \ref{fig:Relics}.
Solid lines show the evolution of $Y_{M}$ for $r_1 = 0.1$ (left) and $r_1 = 0.7$ (right) for fixed values of DM masses $m_1 = 5$ MeV and $m_2 = 30$ MeV, while dashed lines denote equilibrium abundances.
Although the heavier particles $\chi_2$ decouple from the thermal equilibrium earlier than the lighter particles $\chi_1$, their relative abundance, parameterized by $r_1 = \Omega_1/ \Omega_{\text{DM}}$, is determined by the interaction rates $\langle \sigma v \rangle_{22\rightarrow 11}$ and $\langle \sigma v \rangle_{11\rightarrow XX}$.
When $\langle \sigma v \rangle_{22\rightarrow 11} > \langle \sigma v \rangle_{11\rightarrow XX}$,
heavy $\chi_2$ particles are kept in the thermal equilibrium longer, and hence their freeze-out is delayed.
In this case, the $\chi_1$ abundance dominates over $\chi_2$, resulting in a larger fraction of $r_1$.
On the other hand, when $\langle \sigma v \rangle_{22\rightarrow 11} < \langle \sigma v \rangle_{11\rightarrow XX}$,
the freeze-out of $\chi_1$ particles is delayed, so that the $\chi_2$ abundance dominates over $\chi_1$, resulting in a smaller fraction of $r_1$.

Due to the $\chi_1$ annihilation into relativistic SM particles $\langle \sigma v \rangle_{11\rightarrow XX}$, various cosmological data can impose strong constraints on the model~\cite{Kamada:2021muh}. 
The most stringent constraint comes from the CMB where
helium and hydrogen atoms can be ionized by the energy injection from the $\chi_1$ annihilation. 
This can modify the ionization history of electrons after the epoch of the recombination, and therefore disturb the CMB spectra.
In this case, $s$-wave annihilation of $\chi_1$ is strongly disfavored%
\footnote{We expand the cross section in terms of the thermally-averaged relative velocity $\langle \sigma v \rangle \simeq  (\sigma v)_s +  (\sigma v)_p \langle  v^2 \rangle $ where the first and second terms refer to $s$-wave and $p$-wave cross sections respectively.}. 
One possible remedy to this problem is to consider the velocity-suppressed $p$-wave $\chi_1$ annihilation in which case the tension with the CMB can be relieved~\cite{Kamada:2021muh}.
In this work, we will consider the $p$-wave $\chi_1$ annihilation, and assume that $X$ denotes electrons only, while the $\chi_2$ annihilation is fixed to be $s$-wave for simplicity.
Further investigation on the evolution of energy densities of SM particles in Eq.(\ref{eq:BE03}) extends beyond the scope of our analysis, and represents an intriguing topic for future research.

\subsection{Temperature Evolution}
\label{sec:temperature}

Upon the decoupling of DM particles from the SM plasma in temperature $T$, the $\bar{\chi}_2 \chi_2 \rightarrow \bar{\chi}_1 \chi_1$ annihilation can produce $\chi_1$ particles with significant energy, by taking advantage of the mass gap $\delta m = m_2 - m_1$.
Should the interaction rate of the self-scattering $\sigma_{\text{self}1}$ surpasses the Hubble parameter $H$, 
the surplus of kinetic energy is thermally redistributed, raising the overall temperature ($T_1$) of $\chi_1$ particles. In this case, the evolution of temperature $T_1$ is governed by \cite{Kamada:2021muh}
\begin{eqnarray}  
\dot{T}_1  \simeq - 2 H T_1 + \gamma_{\text{heat}} T - 2 \gamma_{\chi_1 X} (T_1 - T) \;,
\label{eq:TempEq}
\end{eqnarray} 
where the first term on the right-hand side represents the Hubble friction,
and the second term accounts for the self-heating,
with $ \gamma_{\text{heat}} $ defined by
\begin{eqnarray}  
\gamma_{\text{heat}} = \frac{2 n^2_2 \langle \sigma v \rangle_{22\rightarrow 11} \delta m }{3 n_1 T} \;.
\label{eq:heat}
\end{eqnarray} 
It should be noted that while a greater mass difference $\delta m$ generally results in a higher temperature $T_1$, it is closely tied to the relative number densities of $\chi_1$ and $\chi_2$ as well.
The last term on the right-hand side of Eq.~(\ref{eq:TempEq}) 
refers to the energy transfer between $\chi_1$ particles and the SM plasma with
\begin{eqnarray}  
\gamma_{\chi_1 X} \simeq \left( \frac{\delta E}{ T}\right ) n_X \langle \sigma v \rangle_{\chi_1 X}, 
\end{eqnarray} 
where $\delta E$ stands for the change in $\chi_1$ kinetic energy per elastic scattering, while $\langle \sigma v \rangle_{\chi_1 X}$ represents the thermally averaged scattering cross section between $\chi_1$ and SM particles \cite{Kamada:2021muh}.

\begin{figure}[t]
    \centering
    \includegraphics[width=0.49\linewidth,clip]{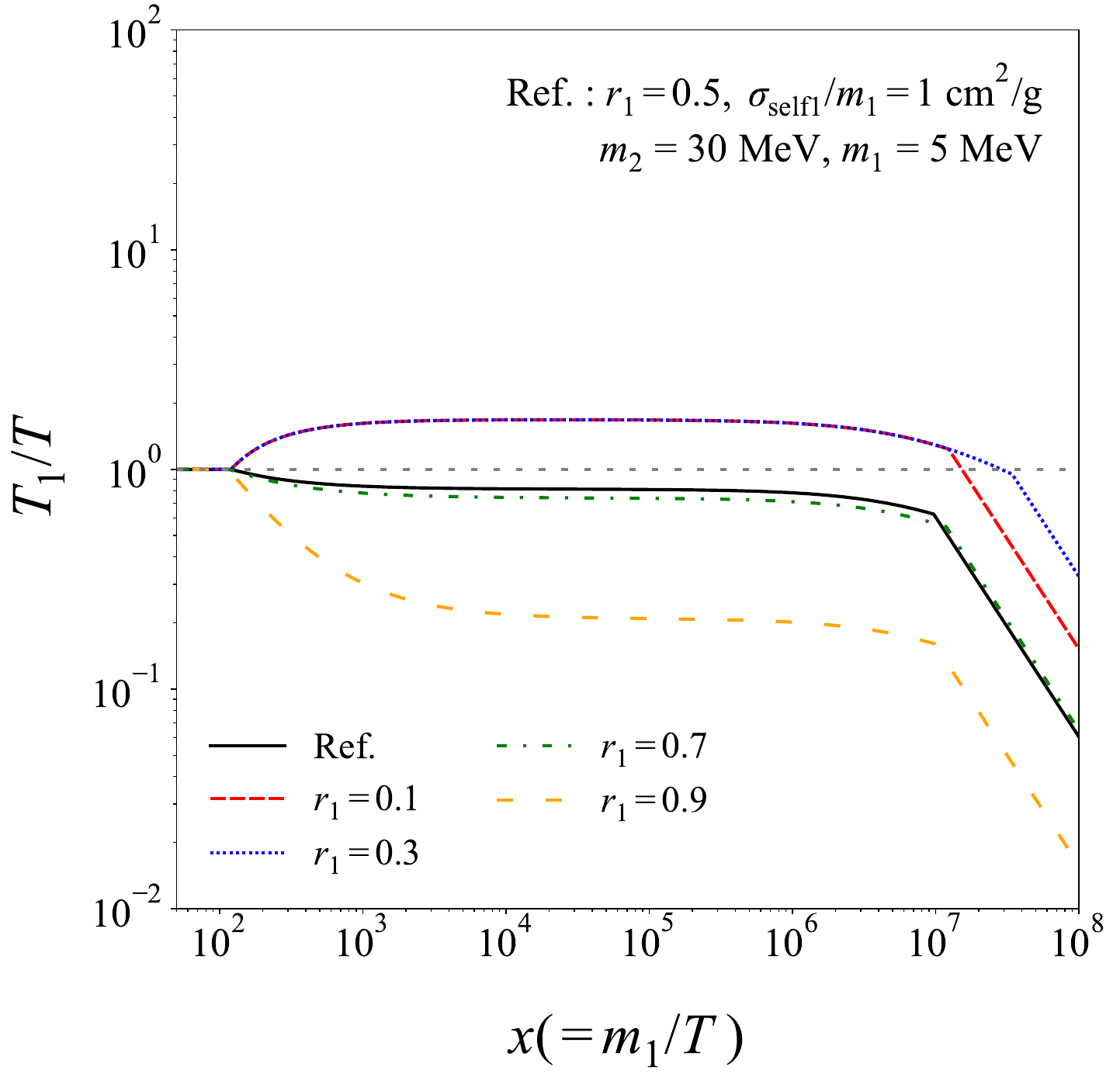}
    \includegraphics[width=0.49\linewidth,clip]{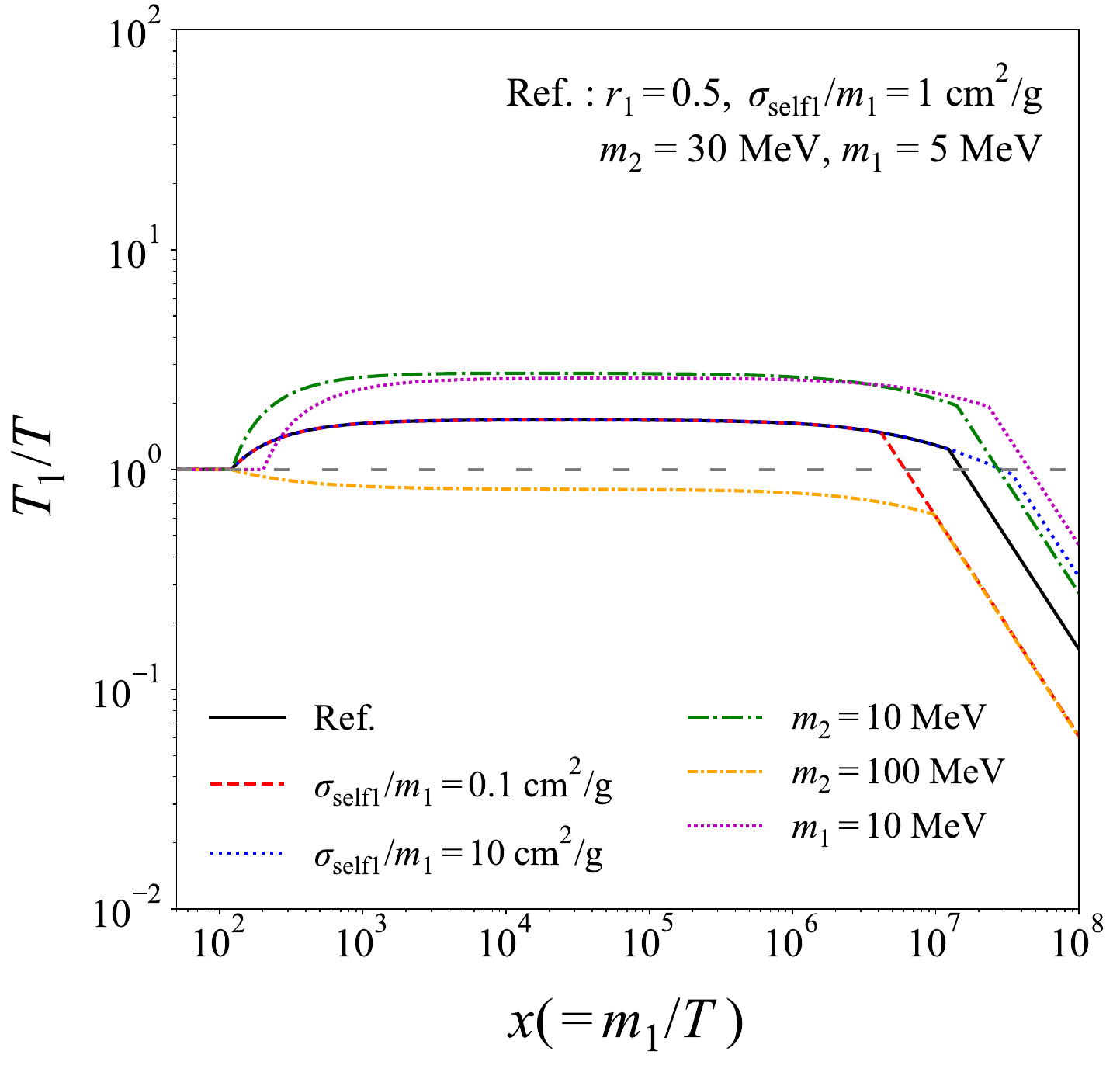}
    \caption{The $\chi_1$ temperature evolutions $T_1$ as a function of $x=m_{1}/T$ under various parameter sets. $T$ denotes the temperature of the radiation bath.    }
    \label{fig:Temperature}
\end{figure}

In our examination of temperature evolution in Eq.~(\ref{eq:TempEq}), we consider three time intervals for practical reasons. 
(i) First, at early times, $\gamma_{\chi_1 X}$ dominates over $\gamma_{\text{heat}}$ and $H$ which ensures that all particles are kept to the kinetic equilibrium with the same temperature $T$. 
In this interval, we simply regard $T_1 = T$.
(ii) Second, as the Universe cools down, when $\gamma_{\chi_1 X}$ drops below the Hubble scale $H$, we turn on the self-heating term $\gamma_{\text{heat}}$.
The temperature ratio $T_1/T$ begins to rise, until it reaches a plateau.
This plateau temperature can be estimated by \cite{Kamada:2021muh}
\begin{equation}
    \Big( \frac{T_1}{T} \Big)_{\text{plat}}  \simeq \frac{\gamma_{\text{heat}}}{H}   
   \simeq 
    \begin{cases}
       \frac{2(1-r_1)}{3 r_1} \frac{m_1 \delta m}{m_2 T^{\infty}_{2} } \Big(\frac{g_* (T^{\infty}_{2}) }{g_* (T_{\text{plat}})} \Big)^{1/2}  
        \cdot \frac{g_{*S} ( T_{\text{plat}} ) }{ g_{*S} ( T^{\infty}_{2} ) }\,, &  (\text{for }T > T_{\text{eq}})  \\
       \frac{4(1-r_1)}{3 r_1} \frac{m_1 \delta m}{m_2 T^{\infty}_{2} }  \Big(\frac{g_* (T^{\infty}_{2}) }{g_* (T_{\text{plat}})} \Big)^{1/2} 
       \cdot \frac{g_{*S} ( T ) }{ g_{*S} ( T^{\infty}_{2} ) } \Big( \frac{T}{T_{\text{eq}}} \Big)^{1/2}\,, &  (\text{for } T < T_{\text{eq}})
    \end{cases}
\end{equation}%
where $T_{\text{eq}} \simeq 0.74 \text{ eV}$ refers to the SM temperature at the matter-radiation equality, $T^{\infty}_{2}$ denotes the freeze-out temperature of $\chi_2$, and $g_{*S}$ is the effective number of relativistic degrees of freedom in entropy.
(iii) Finally, when the interaction rate of the self-scattering $\sigma_{\text{self}1}$ becomes smaller than $H$, we turn off the both $\gamma_{\text{heat}}$ and $\gamma_{\chi_1, X}$ terms from the equation. 
In this interval, $T_1$ simply drops as $T_1 \simeq 1/a^2$. 
The self-scattering effectively terminates when the SM temperature becomes
\begin{eqnarray}  
T_{\text{drop}} \simeq 1 \text{eV} \Big( \frac{T_1}{T} \Big)^{-n}_{\text{plat}} \Big( \frac{0.3}{r_1} \Big)^{2n} \Big( \frac{m_1}{ 100 \text{MeV}} \Big)^n \Big( \frac{1 \text{cm}^2/\text{g}}{\sigma_{\text{self}1}/m_1} \Big)^{2n}
\label{eq:Tdec}
\end{eqnarray} 
where $n = 1/3$ when $T_{\text{drop}} > T_{\text{eq}}$, and $n = 2/9$ when $T_{\text{drop}} < T_{\text{eq}}$.

Figure \ref{fig:Temperature} illustrates how $T_{1}/T$ evolves as a function of $x = m_1/T$ for various parameter benchmarks with each altering one parameter at a time with respect to a reference point
($r_1=0.5$, $\sigma_{\text{self}1}/m_1 = 1~\text{cm}^2/\text{g}$, $m_2 = 30$ MeV, $m_1 = 5$ MeV). 
In the left panel, while keeping all other parameters at their reference values, we vary $r_1$.
To raise the temperature $T_1$, a larger amount of $\chi_2$ density is preferred ($r_1 \ll 1$).
In contrast, if the $\chi_2$ density is smaller than that of $\chi_1$ ($r_1 \to 1$), the amount of energetic $\chi_1$ particles produced from the annihilation of $\chi_2$ will be diminished.
Due to the limited excess energy, it is insufficient to warm a significant portion of the relic $\chi_1$ population, leading to a decrease in the temperature $T_1$.
In the right panel, by changing the self-interaction $\sigma_{\text{self}1}/m_1$ and masses of DM particles, we can observe variations in the $T_{1}/T$ evolution. 
There are a couple of key features that should be noted.
First, the duration of the self-heating effect is determined by $\sigma_{\text{self}1}/m_1$, as expressed in Eq.~(\ref{eq:Tdec}).
As the value of $\sigma_{\text{self}1}/m_1$ rises from $0.1~\text{cm}^2/\text{g}$ (blue) to $10~\text{cm}^2/\text{g}$ (red), there is a noticeable delay in the temperature drop of $T_1$ due to longer heating time.
Second, the combination of number density $n^2_2/n_1$ has a stronger impact on self-heating than the mass difference $\delta m$.
For example, let us fix $m_1 = 5$ MeV, and increase $m_2$ from 30 MeV (black) to 100 MeV (yellow). 
Despite the large $\delta m$, the number density $n_2$ decreases as $n_2 \simeq \rho_2 / m_2$ for non-relativistic particles. 
As a consequence, the temperature $T_1$ in the $m_2 = 100$ MeV case is lower than the $m_2 = 30$ MeV case.
On the other hand, if we fix $m_2 = 30$ MeV and increase $m_1$ from 5 MeV (black) to 10 MeV (purple), the temperature $T_1$ rises due to a decrease in number density $n_1$.

\subsection{Perturbation Evolution}
\label{sec:pertubation}

The impact of DM self-heating on $T_1$ can hinder small-scale astrophysical structures and imprint a residual oscillatory pattern in the matter power spectrum.
To investigate the perturbation evolution of two-component DM, we perturb the 
Boltzmann equations in Eqs.~(\ref{eq:BE01}--\ref{eq:BE03}) following a similar method as discussed in Refs.~\cite{Erickcek:2011us, Barenboim:2013gya, Fan:2014zua, Erickcek:2015jza, Choi:2015yma, Redmond:2018xty}.

We start with the energy-momentum tensors of individual fluids (with $M = 1, 2, X$)
\begin{eqnarray}  
T^{\mu \nu}_{M} = (\rho_{M} + P_{M}) u^{\mu}_{M} u^{\nu}_{M} + P_{M} g^{\mu \nu} \;,
\end{eqnarray} 
where $\rho_{M}$, $P_{M}$, and $u^{\mu}_{M}$ represent energy densities, pressures, and four-velocity vectors respectively, 
and $g^{\mu \nu}$ is a spacetime metric.
It is worth mentioning that each four-velocity (relative to a comoving observer) is normalized by $g_{\mu \nu} u^{\mu} u^{\nu} = -1$, and it is straightforward to show $T_{\nu \lambda} T^{\lambda \beta} u_{\beta} = u_{\nu} \rho^2$.
The energy exchanges caused by DM annihilations can be expressed in covariant forms
\begin{eqnarray}  
\nabla_{\mu} (T^{\mu\nu}_{2} ) &=& - A^{\nu}   \;, \label{eq:P01} \\
\nabla_{\mu} (T^{\mu\nu}_{1} ) &=& A^{\nu} - B^{\nu}  \;, \label{eq:P02} \\
\nabla_{\mu} (T^{\mu\nu}_{X} ) &=& B^{\nu} \;, \label{eq:P03}
\end{eqnarray} 
where the transfer rates can be obtained from Eqs.~(\ref{eq:BE01}--\ref{eq:BE03})
\begin{eqnarray}  
A^{\nu} &=& \frac{\langle \sigma v \rangle_{22\rightarrow 11}}{m_{2}} \left( \rho^2_2 u^{\nu}_2  - \frac{ \rho^2_{2, \text{eq}} }{\rho^2_{1,\text{eq}} }  \rho^2_1 u^{\nu}_1 \right)     \;,   \\
B^{\nu} &=& \frac{\langle \sigma v \rangle_{11\rightarrow XX}}{m_{1}} \Big( \rho^2_1 - \rho^2_{1,\text{eq}}  \Big) u^{\nu}_1 \;.
\end{eqnarray} 

We write the perturbed spacetime in the Newtonian gauge as
\begin{equation}
    ds^2 = -(1 + 2 \Psi)dt^2 + (1 - 2 \Phi) a(t)^2 \delta_{ij} dx^i dx^j \;,
\label{eq:metric}
\end{equation}
where $a(t)$ is a cosmological scale factor, $\Psi$ and $\Phi$ denote scalar perturbations.
We introduce the matter perturbations of individual fluids as
\begin{eqnarray}
    \rho_M = \bar{\rho}_M (1 + \delta_M) \;,
\label{eq:deltaM}
\end{eqnarray}
where $ \bar{\rho}_M$ are background energy densities, and $\delta_M = \delta \rho_M / \bar{\rho}_M$ are density contrasts.
As long as $\delta_M \ll 1$, our perturbation theory remains valid.
To obtain density contrasts of equilibrium densities, we express them as
\begin{eqnarray}
    \delta_{M,\text{eq}} = \frac{n_{M, \text{eq}}}{ \bar{n}_{M, \text{eq}}  } - 1 \;,
\label{eq:deltaEQ1}
\end{eqnarray}
where background equilibrium number densities in the non-relativistic limit are given by
\begin{eqnarray}
	\bar{n}_{M, \text{eq}} = g_M e^{-m_M/\overline{T}} \left( \frac{m_M \overline{T}}{2 \pi} \right)^{3/2} \;,
\end{eqnarray}
where $g_M$ denote degrees of freedom, and we write the temperature perturbation as
$T = \overline{T} (1 + \delta T)$ with $\overline{T}$ denoting a background temperature.
This set of expressions allows us to rewrite Eq.~(\ref{eq:deltaEQ1}) as
\begin{eqnarray}
	\delta_{M, \text{eq}} = \left( \frac{3}{2} + \frac{m_M}{\overline{T}} \right) \delta T \;.
\label{eq:deltaEQ2}
\end{eqnarray}
By taking into account the energy density of photons, $\rho_{\gamma} = 2 \pi^2 T^4/30$, we can formulate the the density contrast of photons as
\begin{eqnarray}
	\delta_{\gamma} = \frac{\rho_{\gamma}}{\bar{\rho}_\gamma} - 1 = 4 \delta T \;.
\label{eq:deltaPhoton}
\end{eqnarray}
We can combine Eq.~(\ref{eq:deltaEQ2}) and Eq.~(\ref{eq:deltaPhoton}) by eliminating $\delta T$, and obtain
\begin{eqnarray}
	\delta_{M, \text{eq}} = \frac{1}{4}  \left( \frac{3}{2} + \frac{m_M}{\overline{T}} \right) \delta_\gamma \;.
\label{eq:deltaEQ3}
\end{eqnarray}
Next, we write the perturbed four-velocity vectors as
\begin{eqnarray}
    u^{\mu}_M = a \left( ( 1 + \Psi),  v^j_M \right) \;,
\label{eq:fourV}
\end{eqnarray}
and define the velocity divergence fields as
\begin{eqnarray}
    \theta_M = \nabla \cdot \vec{v}_M  =  \partial_j v^j_M  \;.
\label{eq:theta}
\end{eqnarray}

We are now ready to derive the matter perturbation equations.
As discussed in the previous section, we will focus on the perturbations of $\chi_1$ and $\chi_2$ fluids, while investigating the perturbations of SM fluids is left for future works.
Utilizing Eqs.~(\ref{eq:metric}--\ref{eq:theta}) and covariant derivatives of energy-momentum tensors~\cite{baumann_2022}, we collect terms in Eqs.~(\ref{eq:P01}, \ref{eq:P02}) at first-order, which leads to the following continuity and Euler equations: 

\begin{footnotesize}
\begin{eqnarray}
    \frac{d \delta_{2}}{dt} + \frac{\theta_{2}}{a} - 3 \frac{d \Phi}{d t} &=& \frac{\big<\sigma v \big>_{2 2\rightarrow 1 1}}{m_{2} \bar{\rho}_{2}} \Bigg( -\Psi \Big( \bar{\rho}^2_{2} - \frac{\bar{\rho}^2_{2,\text{eq}}}{\bar{\rho}^2_{1,\text{eq}}} \bar{\rho}^2_{1} \Big) - \bar{\rho}^2_{2} \delta_{2} + \frac{\bar{\rho}^2_{2,\text{eq}}}{\bar{\rho}^2_{1,\text{eq}}} \bar{\rho}^2_{1} \Big( 2 \delta_{2, \text{eq}} - \delta_{2} -2 \delta_{1 , \text{eq}} + 2 \delta_{1} \Big) \Bigg) \;,~~~~~~~
\label{eq:Perturbation1}\\
\frac{d \theta_{2}}{dt} + H \theta_{2} + \frac{\nabla^2 \Psi}{a} &=& \frac{\big<\sigma v \big>_{2 2\rightarrow 1 1}}{m_{2} \bar{\rho}_{2}}  \frac{\bar{\rho}^2_{2,\text{eq}}}{\bar{\rho}^2_{1,\text{eq}}} \bar{\rho}^2_{1} 
\Big(\theta_{1} - \theta_{2} \Big) \;, 
\label{eq:Perturbation2}\\
\nonumber
\frac{d \delta_{1}}{dt} + \frac{\theta_{1}}{a} - 3 \frac{d \Phi}{d t} &=& -\frac{\big<\sigma v \big>_{2 2\rightarrow 1 1}}{m_{2} \bar{\rho}_{1}} \Bigg( -\Psi \Big( \bar{\rho}^2_{2} - \frac{\bar{\rho}^2_{2,\text{eq}}}{\bar{\rho}^2_{1,\text{eq}}} \bar{\rho}^2_{1} \Big) - \bar{\rho}^2_{2} ( 2\delta_{2} - \delta_{1} ) + \frac{\bar{\rho}^2_{2,\text{eq}}}{\bar{\rho}^2_{1,\text{eq}}} \bar{\rho}^2_{1} \Big( 2 \delta_{2, \text{eq}} + \delta_{1} -2 \delta_{1 , \text{eq}}  \Big) \Bigg) \\ 
&+&  \frac{\big<\sigma v \big>_{1 1\rightarrow X X}}{m_{1} \bar{\rho}_{1}} \Bigg( -\Psi \Big( \bar{\rho}^2_{1} -  \bar{\rho}^2_{1,\text{eq}} \Big) - \bar{\rho}^2_{1} \delta_{1} +  \bar{\rho}_{1,\text{eq}} \Big( 2 \delta_{1,\text{eq}} - \delta_1 \Big) \Bigg) \;,
\label{eq:Perturbation3} \\
\frac{d \theta_{1}}{dt} + H \theta_{1} + \frac{\nabla^2 \Psi}{a} &+& c^2_{s,1} \frac{\nabla^2 \delta_1}{a} =  \frac{\big<\sigma v \big>_{2 2\rightarrow 1 1}}{m_{2} \bar{\rho}_{1}}  \bar{\rho}^2_{2} \Big(\theta_{2} - \theta_{1} \Big)  \;.
\label{eq:Perturbation4}
\end{eqnarray} 
\end{footnotesize}
The terms at zero order correspond to the background Boltzmann equations in Eqs.~(\ref{eq:BE01}--\ref{eq:BE03}) where we dropped {\it bars} in notations for simplicity.

In the absence of non-gravitational interaction, Eqs.~(\ref{eq:Perturbation1}--\ref{eq:Perturbation4}) reproduce the perturbed Boltzmann equations for the CDMs.
The source terms on the right-hand side of Euler equations in Eqs.~(\ref{eq:Perturbation2}, \ref{eq:Perturbation4}) attribute the momentum transfers between $\chi_2$ and $\chi_1$ particles which are directly proportional to the annihilation rate, $\big<\sigma v \big>_{2 2\rightarrow 1 1}$ and to the difference in $\chi_2$ and $\chi_1$ velocities, $\theta_2 - \theta_1$.
As expected, the annihilation attempts to increase the speed of the $\chi_1$ particles in case of a large $\big<\sigma v \big>_{2 2\rightarrow 1 1}$.

The last term on the left-hand side of Eq.~(\ref{eq:Perturbation4}) represents the pressure of $\chi_1$ particles due to the self-heating effect.
The sound speed of $\chi_1$ fluid is determined by
\begin{eqnarray}
    c^2_{s,1} = \frac{T_1}{m_1} \Big( 1 - \frac{1}{3} \frac{\partial \ln T_1}{\partial \ln a} \Big) \;,
\label{eq:cs}
\end{eqnarray}
where the temperature $T_1$ is derived from Eq.~(\ref{eq:TempEq}).
On the contrary, given that the $\chi_2$ particles behave as pressureless CDM, we disregard the pressure element of $\chi_2$ particle in our study.

The evolution of gravitational potentials $\Psi$ and $\Phi$ is governed by two perturbed Einstein equations.
The first equation that we use is its space-time component,
\begin{eqnarray}
    \frac{d \Phi}{dt} + H \Psi = - 4 \pi G a \sum (\bar{\rho} + \bar{P}) \theta \;,
\label{eq:Einstein0s}
\end{eqnarray}
where the summation includes the contributions from photons, baryons, neutrinos, and DM particles.
In particular, to include a pressure term for $\chi_1$ particles, we use an adiabatic approximation $\bar{P}_1 \simeq c^2_{s,1} \bar{\rho}_1$.
Second, we utilize the space-space component of Einstein equation,
\begin{eqnarray}
    ( \partial_i \partial_j - \frac{1}{3} \delta_{ij} \nabla^2 )  (\Phi - \Psi) = 8 \pi G a^2 \Pi_{ij} \;,
\label{eq:Einsteinss}
\end{eqnarray}
where $\Pi_{ij}$ is an anisotropic stress tensor generated by the quadruple moments of photons and neutrinos.

In general, the coupled perturbation equations in Eqs.~(\ref{eq:Perturbation1}--\ref{eq:Perturbation4}, \ref{eq:Einstein0s}--\ref{eq:Einsteinss}) do not have analytic solutions, but we can make progress by conducting our analysis in a special limit. 
The detailed derivation process is provided in Appendix \ref{sec:appendixA}.

\subsection{Linear Power Spectrum}
\label{sec:linear}

Taking into account the complete set of coupled terms and gravitational potentials, we solve the full perturbation equations in Eqs.(\ref{eq:Perturbation1}-\ref{eq:Perturbation4}, \ref{eq:Einstein0s}-\ref{eq:Einsteinss}) numerically.
We have implemented the above perturbation equations and
Boltzmann Eqs.(\ref{eq:BE01}-\ref{eq:BE02}) inside the Cosmic Linear Anisotropy Solving System, \CLASS~(v3.2)~\cite{Blas:2011rf}, which computes the evolution of linear perturbations\footnote{See Refs. \cite{Cyr-Racine:2015ihg, Bansal:2021dfh} for analogous strategy.}.
This implementation is based on an effective theory of structure formation (ETHOS~\cite{Cyr-Racine:2015ihg}), which provides a framework that encapsulates a broad category of dark matter models. Cosmological signature of these models can be computed using only a handful of effective parameters.
We use the following mapping to compute the background and perturbation evolution of $\chi_1$ and $\chi_2$:
\begin{equation}
	\chi_2 \longleftrightarrow \text{CDM}, 
	\quad\quad\quad
	\chi_1 \longleftrightarrow \text{IDM}, 
\end{equation}
where IDM denotes an interacting dark matter species of the {\sc ETHOS} framework.
We turn off all irrelevant interactions of the code.
The temperature evolution of $\chi_1$ follows the procedures outlined in Eq.(\ref{eq:TempEq}) and discussion in secton \ref{sec:temperature}. On the other hand, the $\chi_2$ temperature is not considered.
We modify the two perturbed Einstein equations as shown in Eqs.(\ref{eq:Einstein0s}, \ref{eq:Einsteinss}).
The evolution of energy densities and perturbations of photons, neutrinos, and baryons is analyzed using the standard implementation of \CLASS.
For initial inputs, we use the Planck~2018~\cite{Planck:2018vyg} $\Lambda$CDM parameters,
$\{\Omega_{\rm cdm}h^2,\Omega_b h^2, h,  10^9A_s, n_s, \tau_{\rm reio}\} =\{0.1193, 0.0224, 0.6766,  2.105, 0.9665, 0.0561  \}$.
In solving the scalar perturbation with adiabatic initial conditions, we normalized the initial curvature perturbations to 1.

\begin{figure}[t!]
    \centering
    \includegraphics[width=0.49\linewidth,clip]{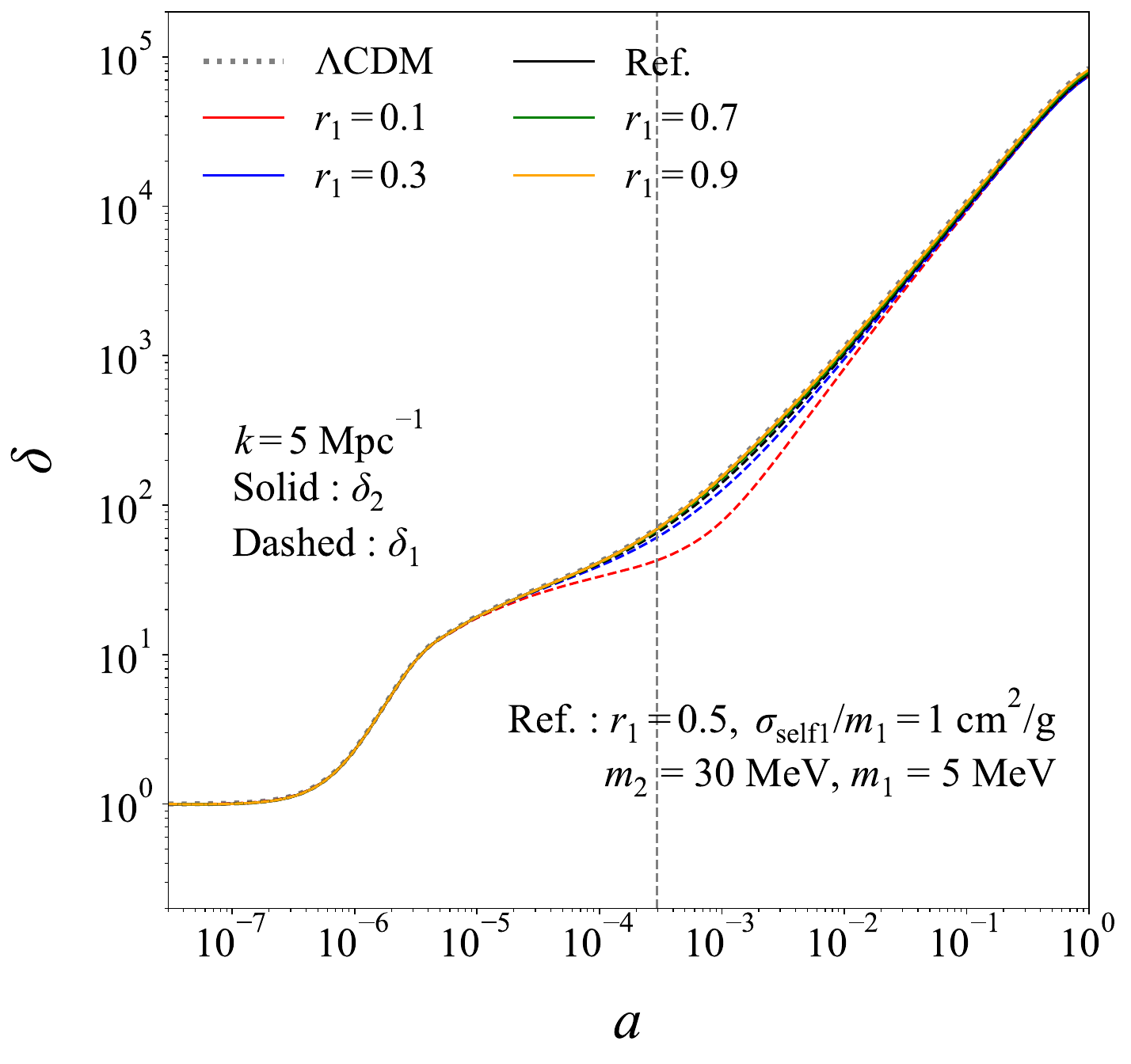}
    \includegraphics[width=0.49\linewidth,clip]{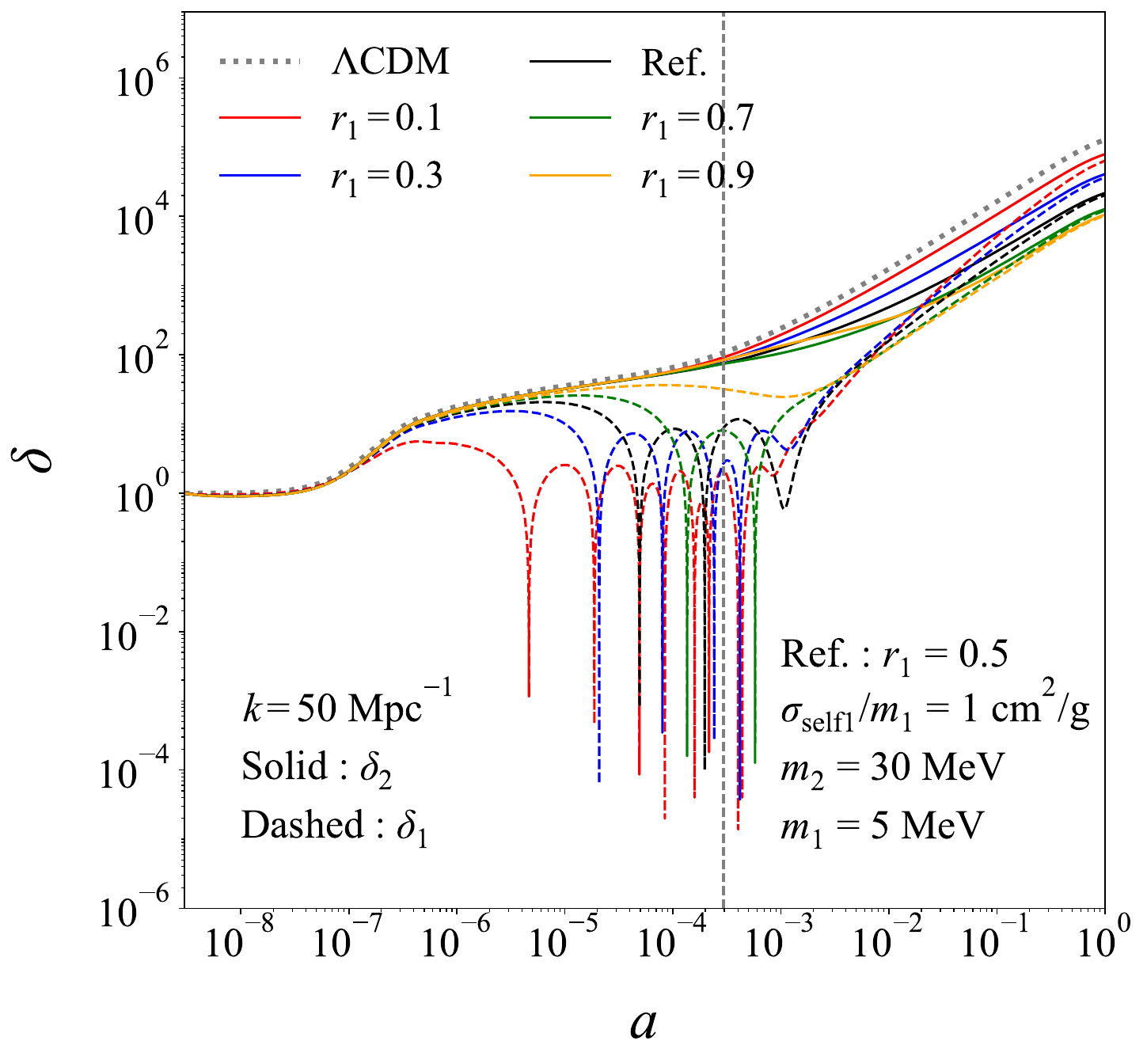}
    \caption{The evolution of $\chi_1$ ($\delta_1$, dashed lines) and $\chi_2$ ($\delta_2$, solid lines) densities as a function of a scale factor $a$ for two fixed modes $k = 5 ~{\rm Mpc}^{-1}$ (left) and $k = 50 ~{\rm Mpc}^{-1}$ (right). 
    Results are shown for different values of $r_1$ for fixed values of $m_1 = 5$ MeV, $m_2 = 30$ MeV, and $\sigma_{\text{self}1}/m_1 = 1 ~\text{cm}^2/\text{g}$.
    The dashed vertical line represents the radiation-matter equality.   
   }
    \label{fig:deltaChi}
\end{figure}

Figure \ref{fig:deltaChi} shows the growth of density contrasts, $\delta_1$ (dashed lines) and $\delta_2$ (solid lines), as a function of a scale factor $a$ for fixed modes $k = 5~ {\rm Mpc}^{-1}$ (left) and $k = 50 ~{\rm Mpc}^{-1}$ (right). 
We vary the ratio $r_1 = 0.1 \sim 0.9$, fixing other parameters $m_1 = 5$ MeV, $m_2 = 30$ MeV, and $\sigma_{\text{self}1}/m_1 = 1~\text{cm}^2/\text{g}$. 
For the lower mode $k = 5~ {\rm Mpc}^{-1}$, the evolution of $\delta_1$ and $\delta_2$ aligns closely with the predictions of the $\Lambda$CDM model (gray dotted line). This suggests that, at this scale, the properties of $\chi_1$ and $\chi_2$ particles do not significantly deviate from the CDM.

For the higher mode $k = 50 ~{\rm Mpc}^{-1}$, on the other hand, both $\delta_1$ and $\delta_2$ exhibit behaviors that starkly different from the $\Lambda$CDM prediction. For the $\delta_1$, an oscillatory pattern emerges just like the SM baryons, inhibiting the growth of matter perturbations. This phenomenon will be termed as Dark Matter Acoustic Oscillations (DMAO)
\footnote{See Ref. \cite{Schaeffer:2021qwm} for a similar effect in a different dark matter model.}.
The DMAO signifies the interplay between the  gravitational attraction and the pressure exerted by the self-heating effect.
In the limit $r_1 \rightarrow 0$, the temperature $T_1$ becomes high. Consequently, the self-heating effect prevails over gravity, leading to DMAO in the perturbation of $\chi_1$.
However, as $r_1$ increases, the temperature $T_1$ drops so that the gravitational force starts to balance out with the pressure, causing $\delta_1$ to stop oscillating.
On the other hand, the evolution of $\delta_2$ is nearly the same as the CDM in the limit $r_1 \rightarrow 0$ where the matter energy density is dominated by $\chi_2$, namely the single-component limit.
There is a slight suppression in $\delta_2$ with increasing $r_1$, where the annihilation cross section $\big<\sigma v \big>_{2 2\rightarrow 1 1}$ becomes large. In this case, the $\delta_2$ experiences a stronger friction due to the disappearance of the gravitational potential well from the annihilation, and hence its overall growth is suppressed. This behavior is in agreement with our prior analysis that we derived from the approximation.

The presence of DMAO would be observable through two distinct forms: an overall suppression of the matter power spectrum, and an oscillatory pattern for the $k$-modes that are greater than the Jeans scale in Eq.(\ref{eq:kJ}).
The linear matter power spectrum, $P(k) = (2\pi)^3 \langle|\delta_m(k)| \rangle^2$, which takes into account the density contrasts of $\chi_1$, $\chi_2$, and baryons, is determined through the calculations given below
\begin{eqnarray}
P(k) &=& \frac{8 \pi^2 k}{25 \Omega^2_m H^4_0} D(a)^2 T(k)^2 A_s \Big( \frac{k}{k_*} \Big)^{n_s-1} \;,
\label{eq:Standard_Power}
\end{eqnarray}
where $k_*= 0.05~\text{Mpc}^{-1}$ denotes a pivot scale.
The transfer function $T(k)$ elucidates the evolution of perturbations as they pass through horizon crossing and the epoch of radiation-matter equality. It is normalized to the unity at the very large scale, $k\to0$
\begin{eqnarray}
T(k) = \frac{\Phi(k, a_{\rm late})}{\Phi(k \rightarrow 0, a_{\rm late})}\;,
\label{eq:Trans}
\end{eqnarray}
where $a_{\rm late}$ represents a scale factor post the transfer function era.
The growth factor $D(a)$ is computed as
\begin{eqnarray}
D(a) = \frac{5}{2} \Omega_m \frac{H(a)}{H_0} \int^a_0~da' \frac{1}{\big( a' \frac{H(a')}{H_0} \big)^3}\;.
\label{eq:Growth_F}
\end{eqnarray}
The numerical values of the $T(k)$ and $D(a)$ are obtained from \CLASS.

\begin{figure}[t]
    \centering
    \includegraphics[width=0.49\linewidth,clip]{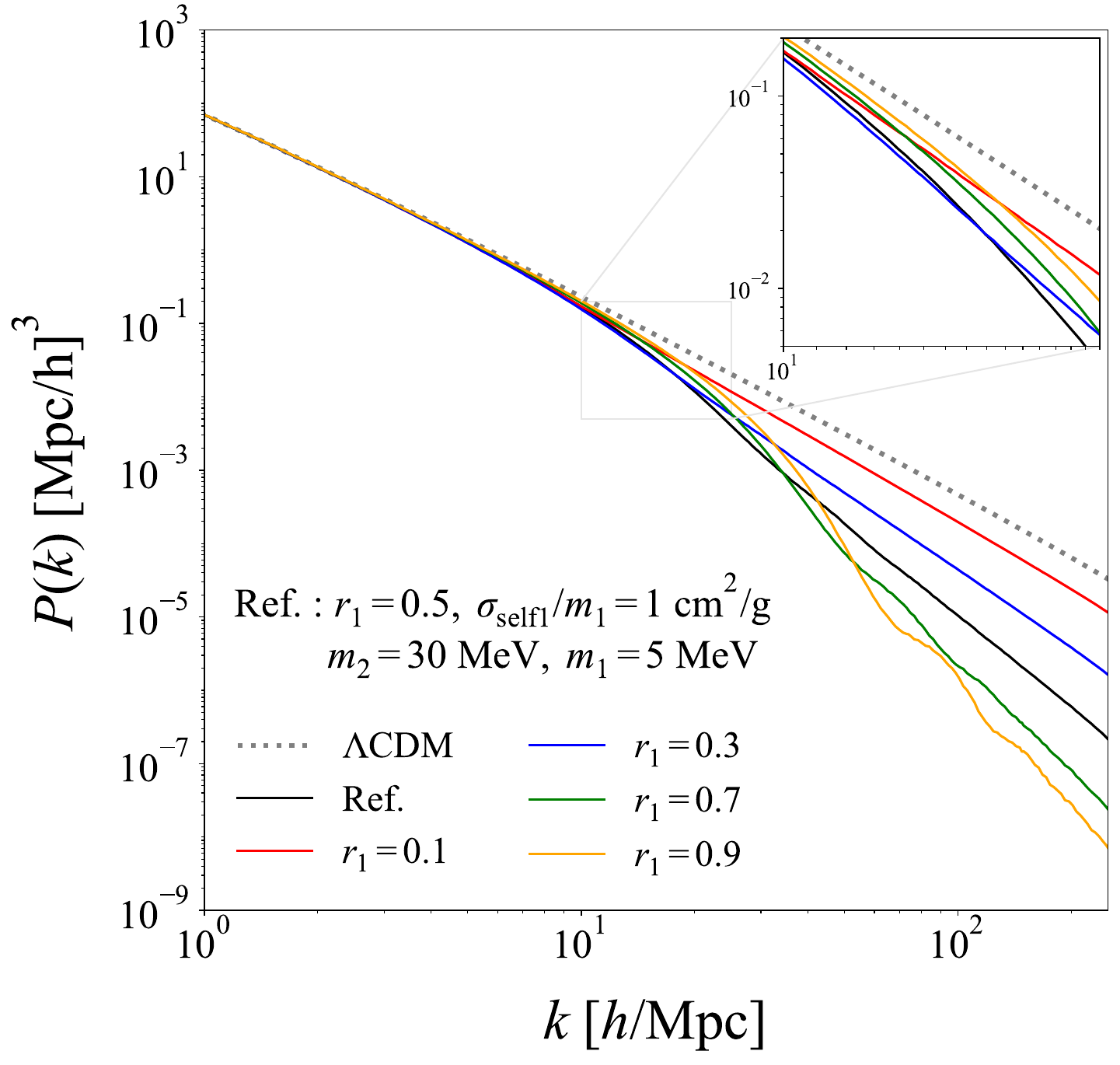}
    \includegraphics[width=0.49\linewidth,clip]{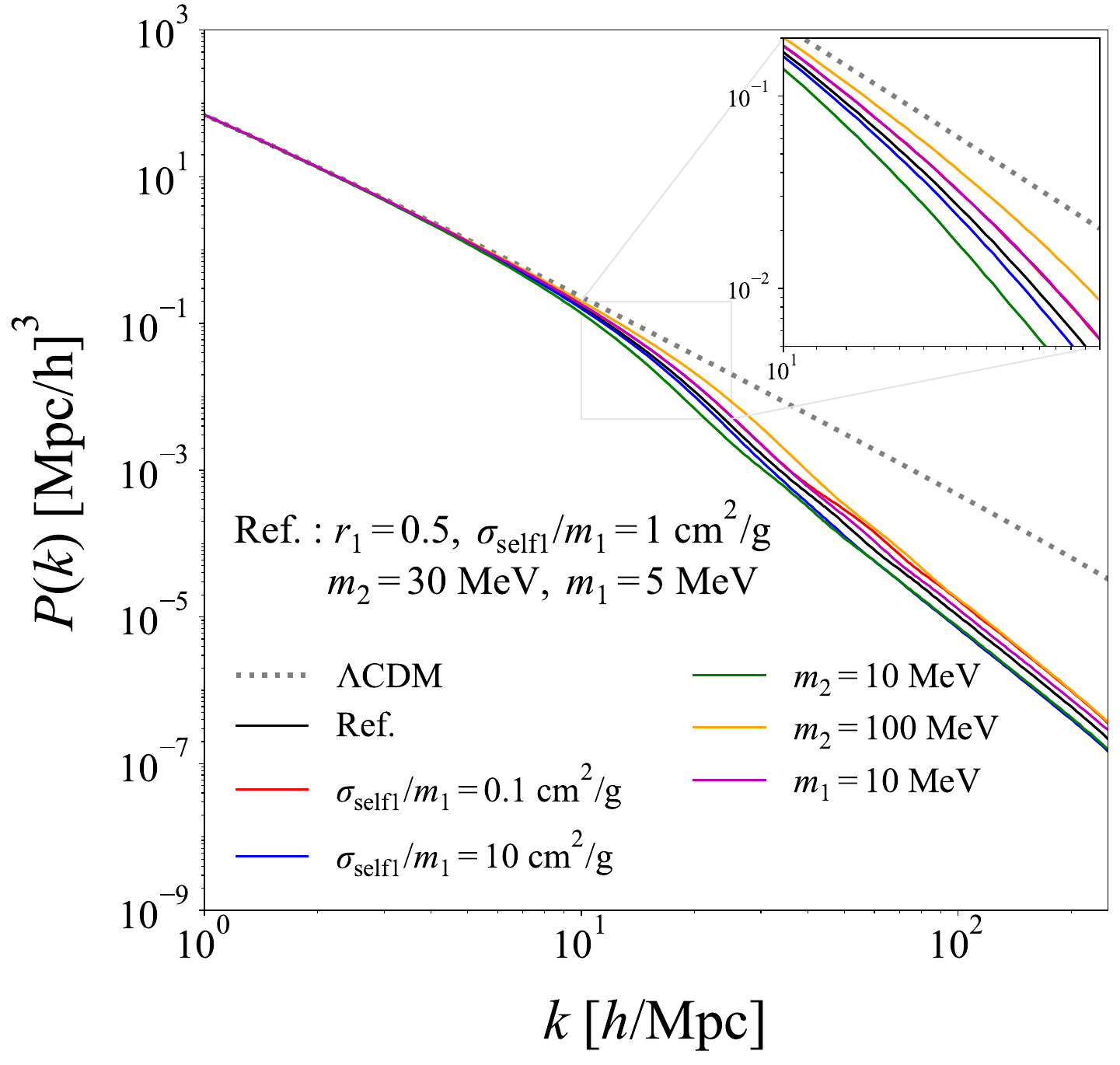}
    \caption{Linear matter power spectra at $z=0$ is calculated using {\tt CLASS} for different values of $r_1$ (left) and  $\sigma_{\text{self}1}/m_1$ or DM masses (right) with respect to the reference parameter set.
    \label{fig:Linear-PK}}
\end{figure}
Figure \ref{fig:Linear-PK} shows the resulting linear matter power spectrum at $z=0$, for various parameter sets, each altering one parameter at a time with respect to a reference point
($r_1=0.5$, $\sigma_{\text{self}1}/m_1 = 1~\text{cm}^2/\text{g}$, $m_2 = 30$ MeV, $m_1 = 5$ MeV). 
In the left panel, we vary $r_1$ while keeping all other parameters at their reference values. When $\chi_2$ accounts for the majority of DM ({\it e.g.,} $r_1$ = 0.1), the power spectrum aligns with the $\Lambda$CDM (gray dotted line) prediction, albeit there is a slight suppression in high $k$-modes.
As we gradually increase $r_1$, the suppression in the power spectrum becomes more pronounced. This behavior is a reminiscent of WDM which washes out the scale below the free-streaming length.
At $k$-modes of $\mathcal{O}(10~h{\rm Mpc^{-1}})$, around Jeans scales, the most significant suppression occurs at $r_1 \sim 0.5$ (black solid line). However, if we continue to increase $r_1$, the power spectrum reverts to the $\Lambda$CDM prediction.
This behavior can be understood as follows. 
In the limit $r_1 \rightarrow 0$, the temperature $T_1$ rises significantly (see Figure \ref{fig:Temperature}), yet the number density of $\chi_1$ is insufficient to generate an impact on the power spectrum.
If we consider the opposite limit $r_1 \rightarrow 1$, although the $\chi_1$ number density becomes higher, the temperature $T_1$ is much lower. This indicates that the self-heating effect has the most profound influence on structure formation in the admixture state at around $r_1 \sim 0.5$.
This phenomenon will alter for higher modes $k \gtrsim 100~h{\rm Mpc^{-1}}$, where the largest suppression in the power spectrum occurs at $r_1 \sim 1$. This is because the pressure term, $c^2_{s,1} \frac{k^2}{a^2} \delta_1$, in Eq.(\ref{eq:2ndDelta2}) grows as $k$ increases.
Although $T_1$ temperature declines as $r_1 \rightarrow 1$ and hence the sound speed $c^2_{s,1}$ is small, this can be compensated by the higher values of $k$, which renders a condition for the DMAO.
Therefore, small-scale structures below the scale of $\mathcal{O}(10~h^{-1}{\rm kpc})$ are expected to be dampened even at the larger $r_1$ ratio.
Another important remark is that there are residual oscillatory patterns at higher modes $k \gtrsim 100~h{\rm Mpc^{-1}}$ resulting from the DMAO phenomena. 
Such an oscillatory behavior is a common feature of  two-component DM models%
\footnote{See Refs. \cite{Chacko:2018vss,Bansal:2021dfh} for a similar oscillatory pattern in the matter power spectrum in the Mirror Twin Higgs cosmology.}
that can be potentially observed in future experiments for large scale structure.

The right panel in Figure \ref{fig:Linear-PK} shows the changes in the power spectrum in response to alternations in the self-interaction $\sigma_{\text{self}1}/m_1$ and DM masses.
The power spectrum experiences greater suppression with an increase in $\sigma_{\text{self}1}/m_1$, as it enhances the self-heating effect.
On the other hand, the suppression is lifted for heavier $\chi_2$ (lighter $\chi_1$) particles because its number density becomes smaller (larger), leading to a decrease in $T_1$ (see Eq.(\ref{eq:heat})).

\begin{figure}[t]
    \centering
    \includegraphics[width=0.49\linewidth,clip]{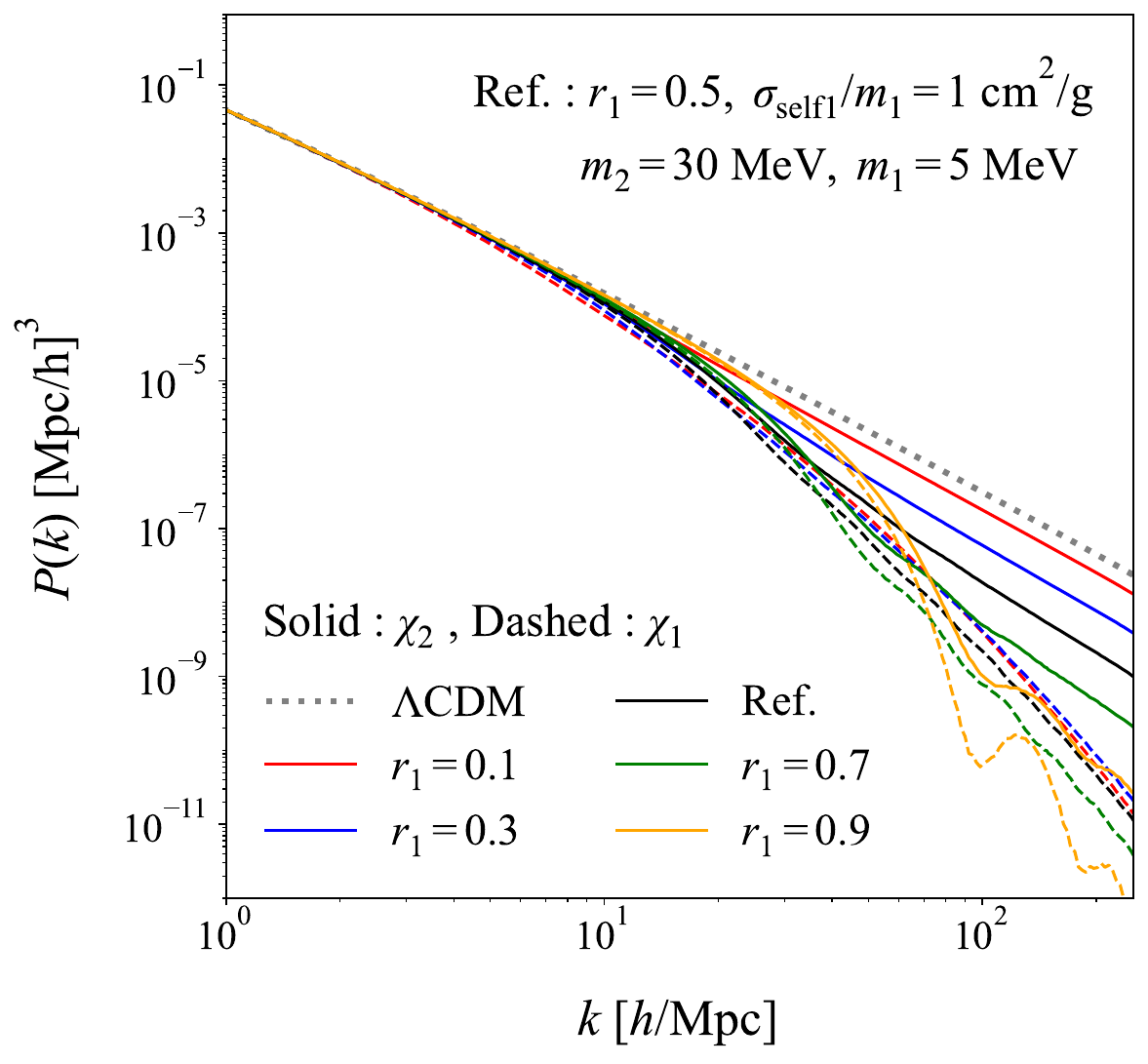}
    \includegraphics[width=0.49\linewidth,clip]{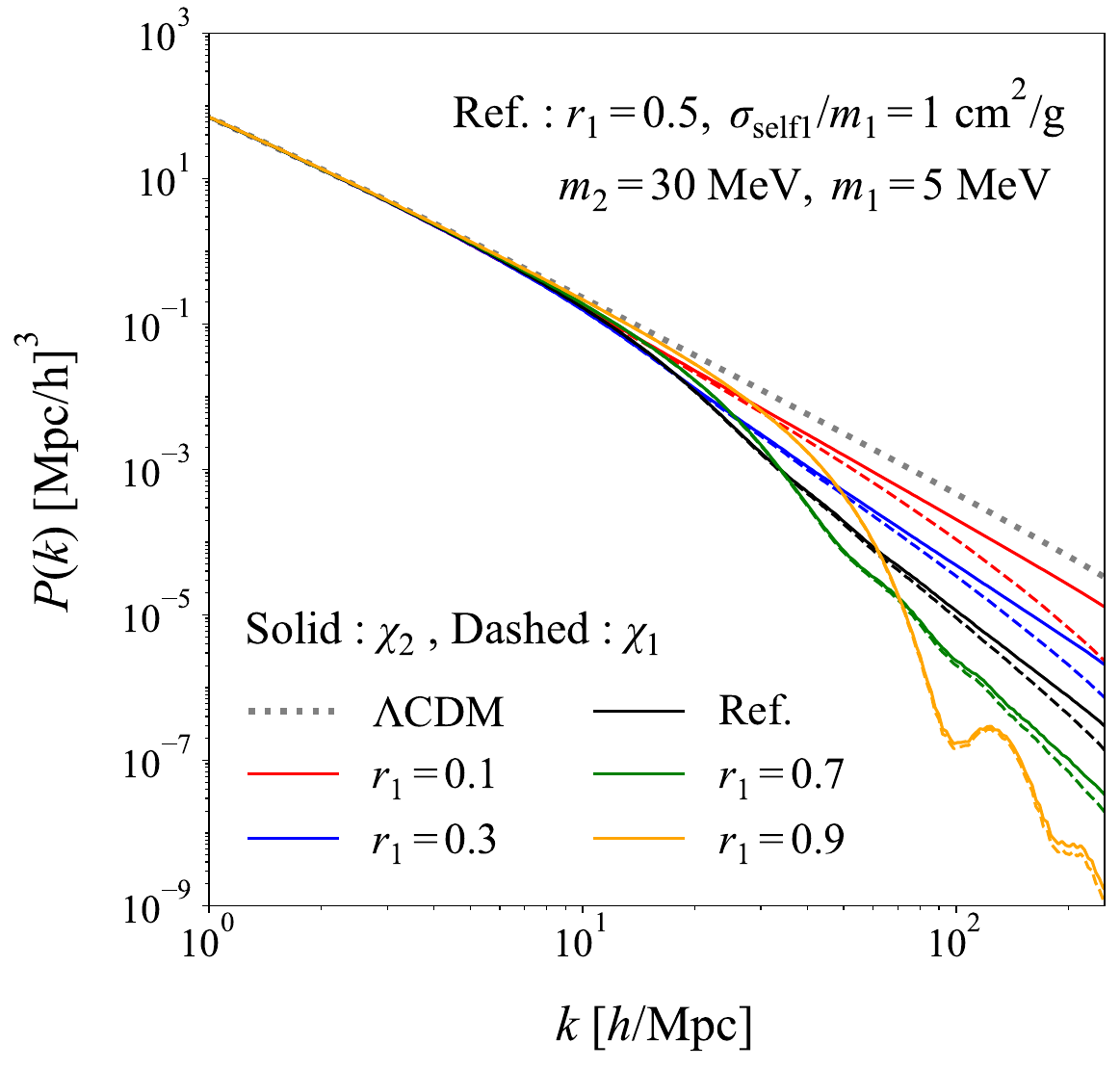}
    \caption{
    Individual power spectra of $\chi_1$ (dahsed lines) and $\chi_2$ (solid lines) at the redshift of $z = 49$ (left) and $z = 0$ (right), for varying $r_1$ with respect to the reference parameter set.     }
    \label{fig:Linear_PK_Sep}
\end{figure}
The linear matter power spectra above will be used to set up $N$-body simulations in Section \ref{sec:nbody}, to incorporate non-linear effects. For this purpose, we break down the matter power spectrum into two 
separate components for $\chi_1$ and $\chi_2$ by%
\footnote{See Ref. \cite{Anderhalden:2012qt} for a similar method of decomposing individual power spectra in the mixed CDM and WDM model.}
\begin{eqnarray}
P_{\chi_1}(k) &\simeq& P_{\Lambda {\rm CDM}} \Big( \frac{\delta_1}{\delta_{\Lambda {\rm CDM}}} \Big)^2 \;, \nonumber  \\
P_{\chi_2}(k) &\simeq& P_{\Lambda {\rm CDM}} \Big( \frac{\delta_2}{\delta_{\Lambda {\rm CDM}}} \Big)^2 \;,
\label{eq:Separated_Power}
\end{eqnarray}
where $\delta_{\Lambda {\rm CDM}}$ and $P_{\Lambda {\rm CDM}}$ denote the density contrast and the power spectrum of CDM obtained from \CLASS~using the same initial $\Lambda$CDM parameters above.
Figure \ref{fig:Linear_PK_Sep} shows the individual power spectra of $\chi_1$ (dahsed lines) and $\chi_2$ (solid lines) at the redshift of $z = 49$ (left panel) and $z = 0$ (right panel), varying $r_1$ with respect to the reference parameter set.

At the redshift of $z = 49$, the suppression of the power spectrum is more pronounced for $\chi_1$ than for $\chi_2$ since the DMAO prohibits the growth of $\delta_1$.
Compared to the CDM (gray dotted line), the $\chi_2$ power spectrum is  suppressed due to the friction caused by the disappearance of the gravitational potential well during the $\chi_2$ annihilation.
Oscillatory patterns can be observed in both power spectra of $\chi_1$ and $\chi_2$. This can be explained by the transmission of DMAO effects from the $\chi_1$ perturbation evolution to the $\chi_2$ sector via gravitational interactions.
As they evolve towards a redshift of $z = 0$, the perturbations of the dominant species start to gravitationally attract the other species, leading to comparable power spectra shapes.

\section{$N$-body Simulation: Impact on Structure Formation}
\label{sec:nbody}

Linear perturbation theory effectively captures the growth of matter density contrast on large scales, but its evolution becomes significantly non-linear on smaller scales with $k \gg 0.1~h{\rm Mpc}^{-1}$. 
Accurate calculation of the non-linear evolution necessitates extensive $N$-body simulations, the outcomes of which can be compared to basic empirical parameterizations.
In this section, we perform $N$-body simulations to include non-linear corrections and analyze various small-scale cosmological structures predicted by the two-component DM model.

The simulation of two-component DM~\cite{Todoroki_2018} is carried out with \texttt{GADGET-3}~\cite{Springel_2005,Springel_2008}.
To provide an initial input, we use the Planck~2018~\cite{Planck:2018vyg} $\Lambda$CDM parameters, \{$\Omega_{\Lambda}$,\,$\Omega_{\rm cdm}$,\,$\Omega_{b}$,\,$h$,\,$\sigma_{8}$\} = \{0.6889, 0.2606, 0.049, 0.6766, 0.8102\}.
The size of the periodic comoving box is set $3~h^{-1}\rm{Mpc}$ with a total of $N_{\rm tot}=128^3$ DM clumps. 
The numbers of $\chi_1$ and $\chi_2$ clumps are set $N_1 = N_{\rm tot} r_1$ and $N_2 = N_{\rm tot} (1 - r_1)$, respectively.
The simulation starts at a redshift of $z_i = 49$ and runs until a redshift of $z_f = 0$.
By utilizing the linear matter power spectra of $\chi_1$ and $\chi_2$ derived from Eq.~(\ref{eq:Separated_Power}) at $z=49$, we set up the initial density perturbation of each species through second-order Lagrangian perturbation theory. 
This is achieved by displacing the positions and velocities of initially-homogeneous DM clumps using the built-in function \texttt{2LPTIC}~\cite{Crocce_2006}\footnote{In $N$-body simulations with WDM~\cite{paduroiu2015structure,Leo_2017,Bode_2001} or mixed CDM-WDM \cite{Maccio:2012rjx, Parimbelli:2021mtp,Boyarsky_2009,Macci__2012,Anderhalden_2012},
the Zel'dovich velocity is adjusted to include the thermal velocity of WDM particles to incorporate their thermal effects. 
Nonetheless, in our simulation, the masses of the $\chi_1$ and $\chi_2$ particles are significantly larger than that of WDM, making thermal effects negligible.
Furthermore, we use a forward method~\cite{Angulo:2021kes} connecting the output of the Einstein-Boltzmann code (\CLASS) at $z_i = 49$ and the simulations.
}.
The gravitational softening length, $\epsilon= 0.001$, determines the force resolution for individual DM clumps. Its main function is to weaken the gravitational force so that DM clumps do not experience intense acceleration from nearby clumps.

In the two-component DM simulation, $\chi_1$ DM clumps can interact non-gravitationally with each other elastically via self-interaction.
The code employs the Monte-Carlo technique to simulate the $\chi_1$ self-interaction. 
The method is commonly used to simulate a typical Self-Interacting Dark Matter (SIDM) in $N$-body simulations~\cite{Rocha:2012jg, Ahn:2004xt, Colin_2002, Moore:2000fp, Yoshida_2000, Spergel:1999mh}.
The probability of the self-interaction taking place within $\Delta t$ is calculated by
\begin{eqnarray}
P_{i j \rightarrow i' j'}~ = ~\rho_j ~ \Big(\frac{\sigma_{\text{self}1}}{m_1} \Big) ~ |\vec{v}_j - \vec{v}_i | ~\Delta t~ \Theta(E_{i'j'}) \;,
\label{eq:Prob}
\end{eqnarray}
where $i$ and $j$ respectively denote a projectile clump and a target clump,
$\rho_j$ is a local density of the target clump\footnote{The local density $\rho_j$ is defined by
\begin{eqnarray}
\rho_j = \sum^{N}_{k=1} m_k W(|\vec{r}_k - \vec{r}_j |, h_j) \;,
\label{eq:rho}
\end{eqnarray}
where $k$ denotes a neighboring $\chi_1$ clump, $W$ is a cubic spline kernel function~\cite{1992ARA&A..30..543M}, $\vec{r}$ is a position of each clump, and $h_j$ represents the adaptive smoothing radius of the target clump, encompassing $N=33$ neighboring $\chi_1$ clumps within a spherical volume.
},
$\sigma_{\text{self}1}/m_1$ is the $\chi_1$ self-scattering cross
section per unit mass, $|\vec{v}_j - \vec{v}_i |$ is a relative velocity of the initial pair, and the Heaviside function, denoted by $\Theta(E_{i'j'})$, eliminates kinematically forbidden configurations of outgoing clumps.
The momenta of outgoing clumps are back-to-back, scattering isotropically.
Mapping the particle physics cross section, $\sigma_{\text{self}1}$, into the interaction between astronomical DM clumps involves intricate subtleties. 
Our assumption here is that the $\chi_1$ self-scattering is isotropic, like in the $s$-wave process, making the cross section independent of a scattering angle.
In this scenario, such a mapping may be considered acceptable in numerical simulations.
For more in-depth discussion, we suggest Ref.~\cite{Tulin:2017ara} and the references therein to dedicated readers.
On the other hand, as mentioned previously, 
in order to focus on the main dynamics of $\chi_1$ self-interaction, we will exclude the influence of $\chi_2$ self-interaction in our simulations. 
Furthermore, given that the annihilation rate of $\chi_2$ into $\chi_1$ particles is much smaller than the Hubble parameter in the late universe, we will not consider the annihilation effect in the simulations.

\subsection{Nonlinear Power Spectrum}
\label{sec:powerspectrum}

%
\begin{figure*}[t]
\centering
\includegraphics[width=1\linewidth,clip]{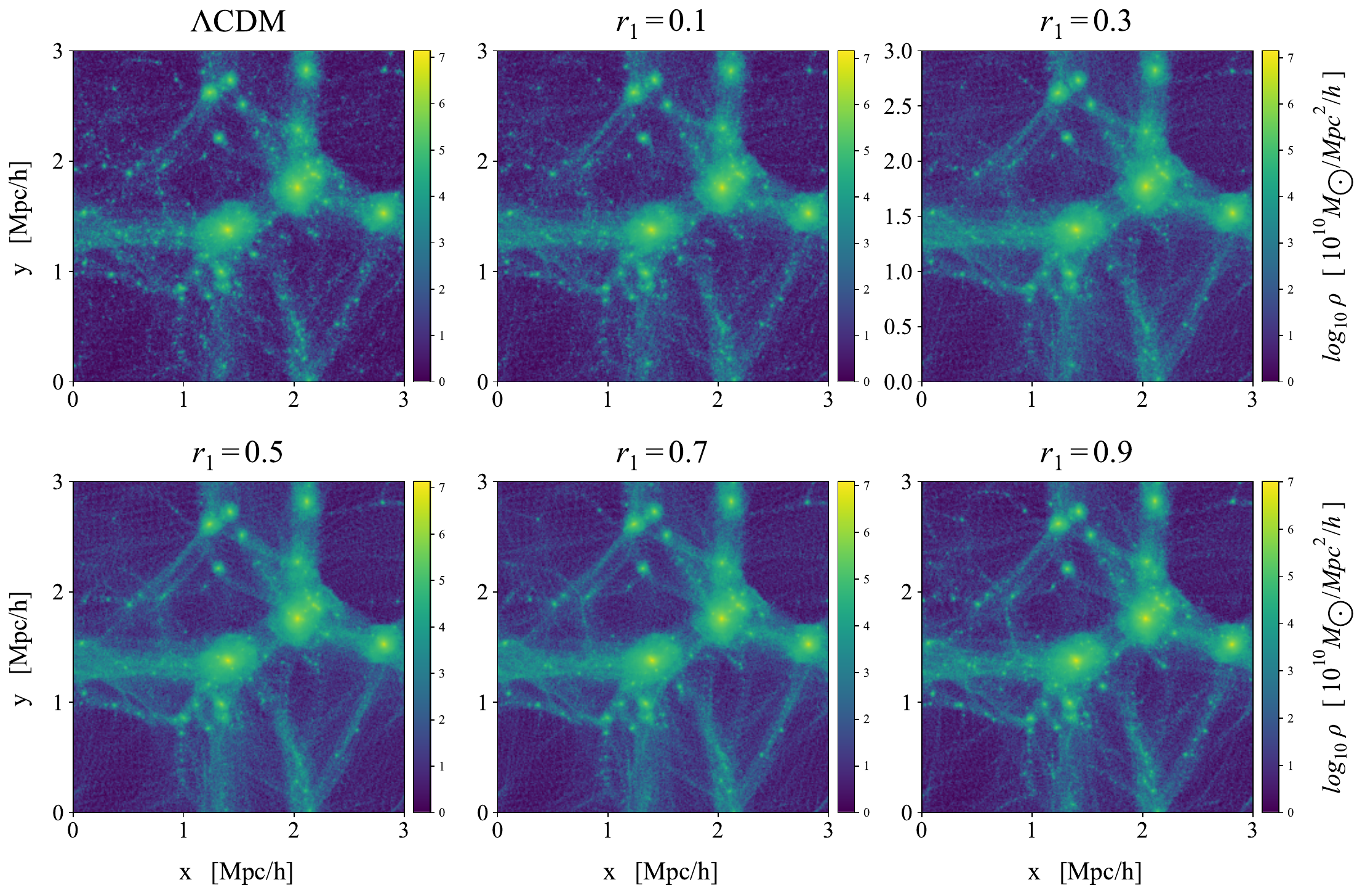}
\caption{Visualization of DM density fields in the periodic $3~h^{-1}\rm{Mpc}$ box at $z=0$, comparing the $\Lambda$CDM simulation (top-left panel) with the two-component DM scenarios with varying $r_1 \in [0.1, 0.9]$ (starting from the top-middle panel in a rightward direction to the bottom).
}
\label{fig:LSS} 
\end{figure*}

Figure~\ref{fig:LSS} illustrates the visualization of DM density fields in the periodic $3~h^{-1}\rm{Mpc}$ box at $z=0$, comparing the $\Lambda$CDM simulation (top-left panel) with the two-component DM scenarios with varying $r_1 \in [0.1, 0.9]$ (starting from the top-middle panel in a rightward direction to the bottom).
When considering the overall statistics of DM, numerous small satellite fields are observable in the $\Lambda$CDM simulation.
On the other hand, as we gradually increase $r_1$ in the two-component DM scenarios, the number of small satellites decreases, with the most notable reduction happening around $r_1 \sim 0.5$.
However, increasing $r_1$ further leads to a slight rise in the number of small satellites, being rather closer to the $\Lambda$CDM prediction.
This is consistent with the behavior of the linear power spectrum in Figure~\ref{fig:Linear-PK} 
at $k$ modes of $\mathcal{O}(50~h{\rm Mpc^{-1}})$ that roughly match the size of the small satellites.

\begin{figure}[t]
    \centering
    \includegraphics[width=0.49\linewidth,clip]{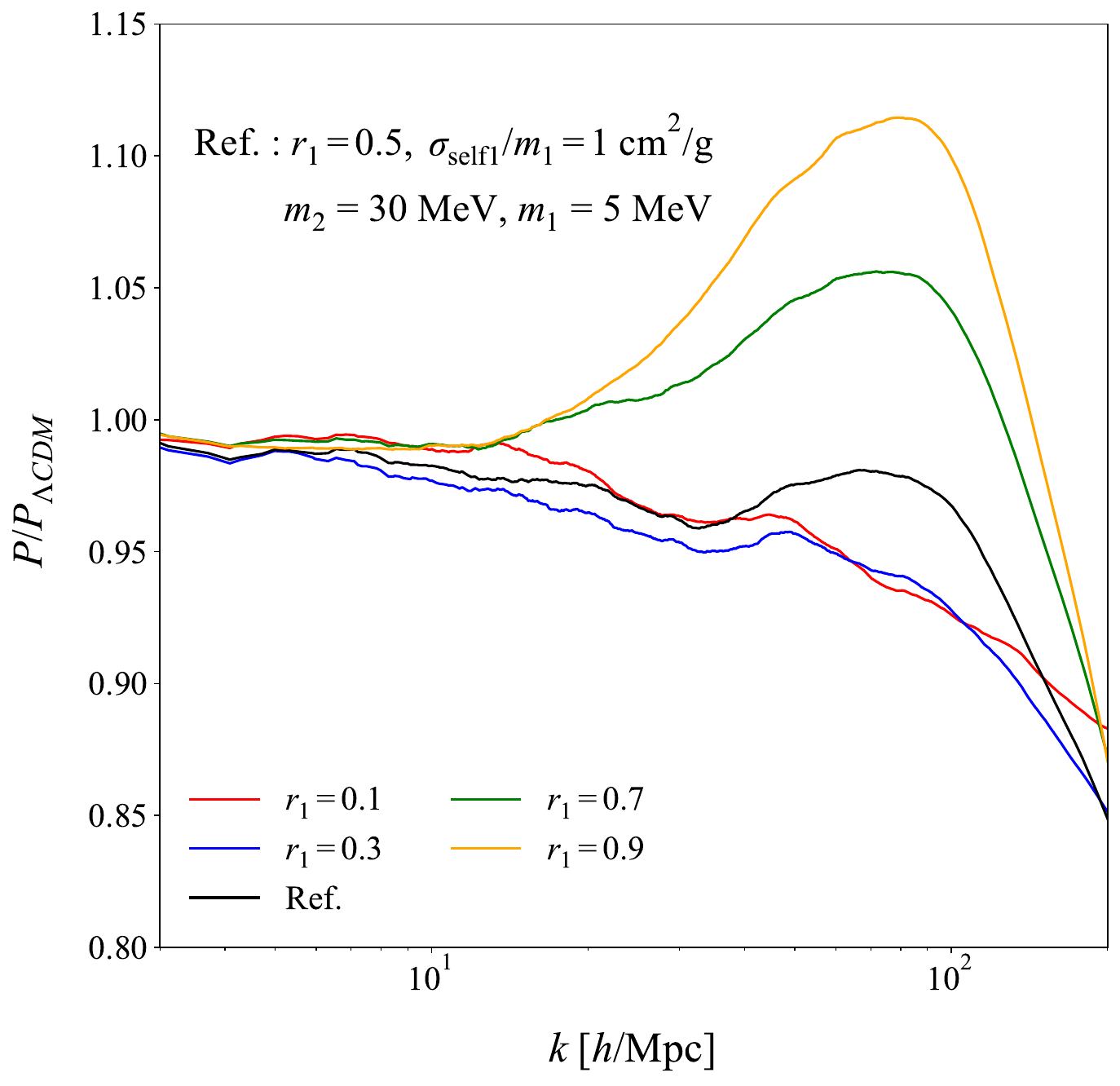}
    \includegraphics[width=0.49\linewidth,clip]{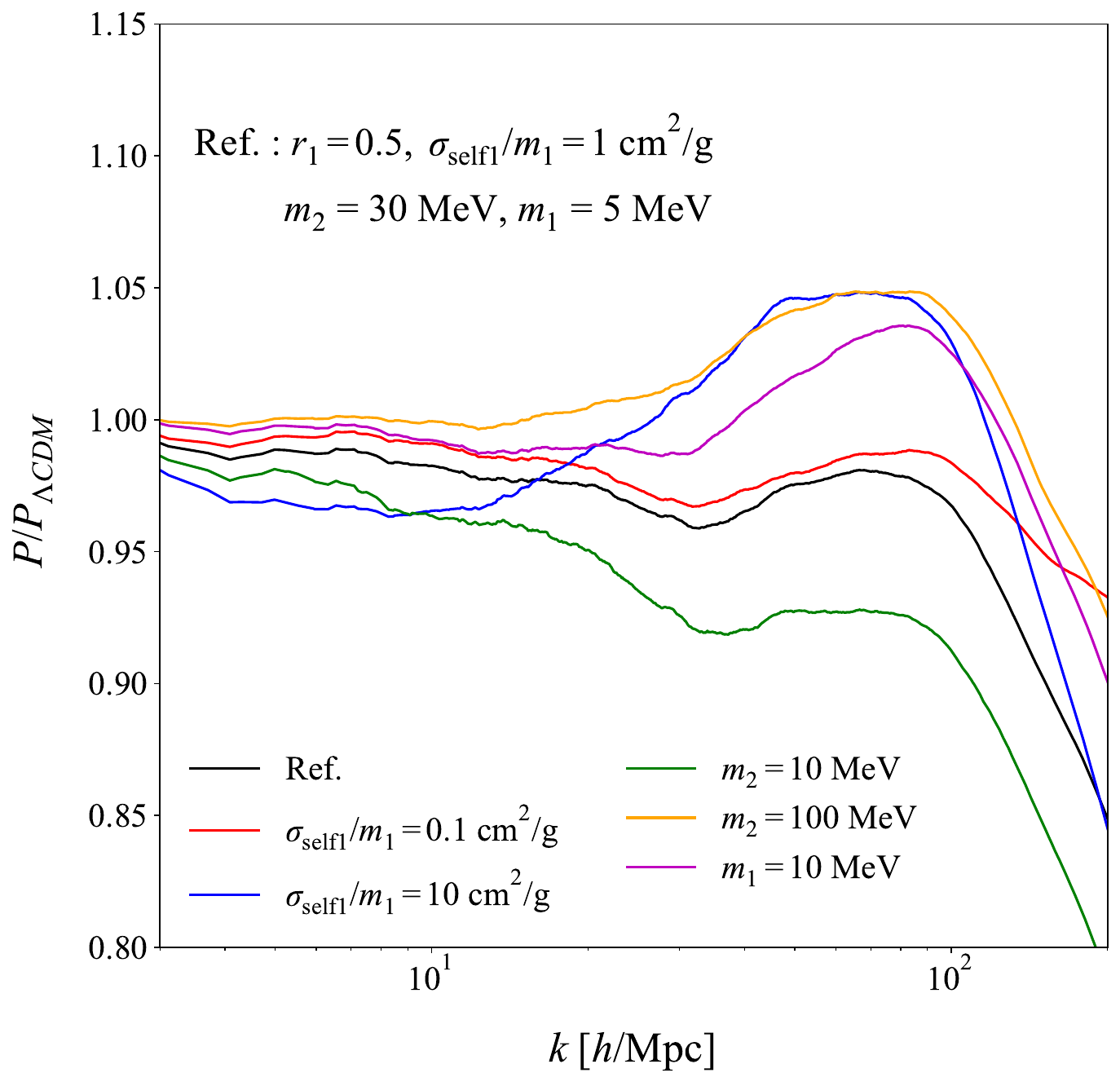}
    \caption{The ratio of the non-linear power spectra between the two-component DM scenario and the $\Lambda$CDM model for varying (left) $r_1$ and (right) $\sigma_{\text{self}1}/m_1$ or DM masses with respect to the reference parameter set.       }
    \label{fig:Non-Linear-Power}
\end{figure}

Figure~\ref{fig:Non-Linear-Power} illustrates the ratio of the non-linear power spectra between the two-component DM scenario and the $\Lambda$CDM model.
When compared to the linear power spectra in Figure~\ref{fig:Linear-PK}, which exhibit notable differences from the $\Lambda$CDM, non-linear effects have a considerable impact on washing out the linear features.
Around $k$-modes of $\mathcal{O}(10~h{\rm Mpc^{-1}})$, there are only deviations $\lesssim 5\%$ from the $\Lambda$CDM.
In this prospect, the Lyman-$\alpha$ observations~\cite{Viel:2013fqw} which can probe the matter power spectrum on scales $ 0.5 < k < 20~h{\rm Mpc^{-1}}$ might not offer compelling bounds on the two-component DM.
The enhancement of the power spectrum at the scales $k \sim 100~h{\rm Mpc^{-1}}$ for higher $r_1$ values can be explained by the $20\%$ higher density of $\chi_1$ in the outer region of a galaxy, $r \sim 10~{\rm kpc}$, compared to the $\Lambda$CDM prediction as shown in Figure \ref{fig:density_sep}. This occurs because of the $\chi_1$ self-interaction, causing the redistribution of $\chi_1$ clumps from the center to the outer region.
In the central region of a galaxy, on the other hand, $\chi_1$ densities are $\sim 70\%$ lower than the $\Lambda$CDM prediction as shown in Figure \ref{fig:density_sep}, which explains the $\mathcal{O}(10\%)$ suppression of the power spectrum at $k \gg 100~h{\rm Mpc^{-1}}$.

\subsection{Sub-halo Structure}
\label{sec:subhalo}

To extract halo properties, one needs to identify groups of DM clumps (referred to as halos) that surpass the virial overdensity, for which we use the Friends-of-Friends (FOF) method~\cite{Springel:2020plp}. 
The SUBFIND algorithm \cite{Springel:2000qu} is adopted to identify substructures (referred to as sub-halos) in those halos. 
The linking length of FOF algorithm is 0.2. The minimum masses of the halos and sub-halos are $3.56 \times 10^{7}{M_{\odot}}$ and $2.22 \times 10^{7}{M_{\odot}}$ respectively. The most massive halos in the simulation are around $\mathcal{O}(10^{11})M_{\odot}$ which is slightly smaller than the Milky Way.

\begin{figure}[t]
    \centering
    \includegraphics[width=0.49\linewidth,clip]{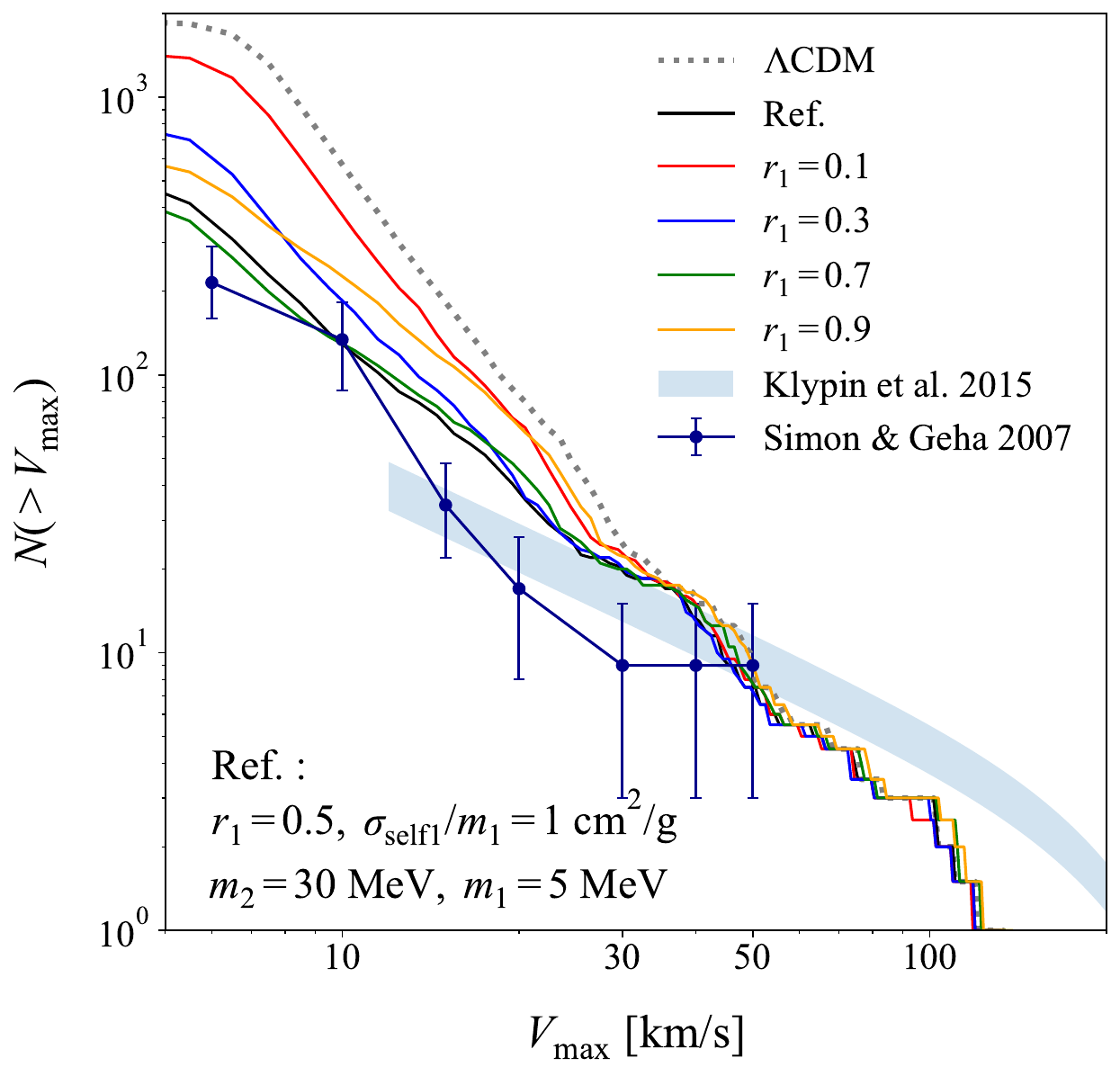}
    \includegraphics[width=0.49\linewidth,clip]{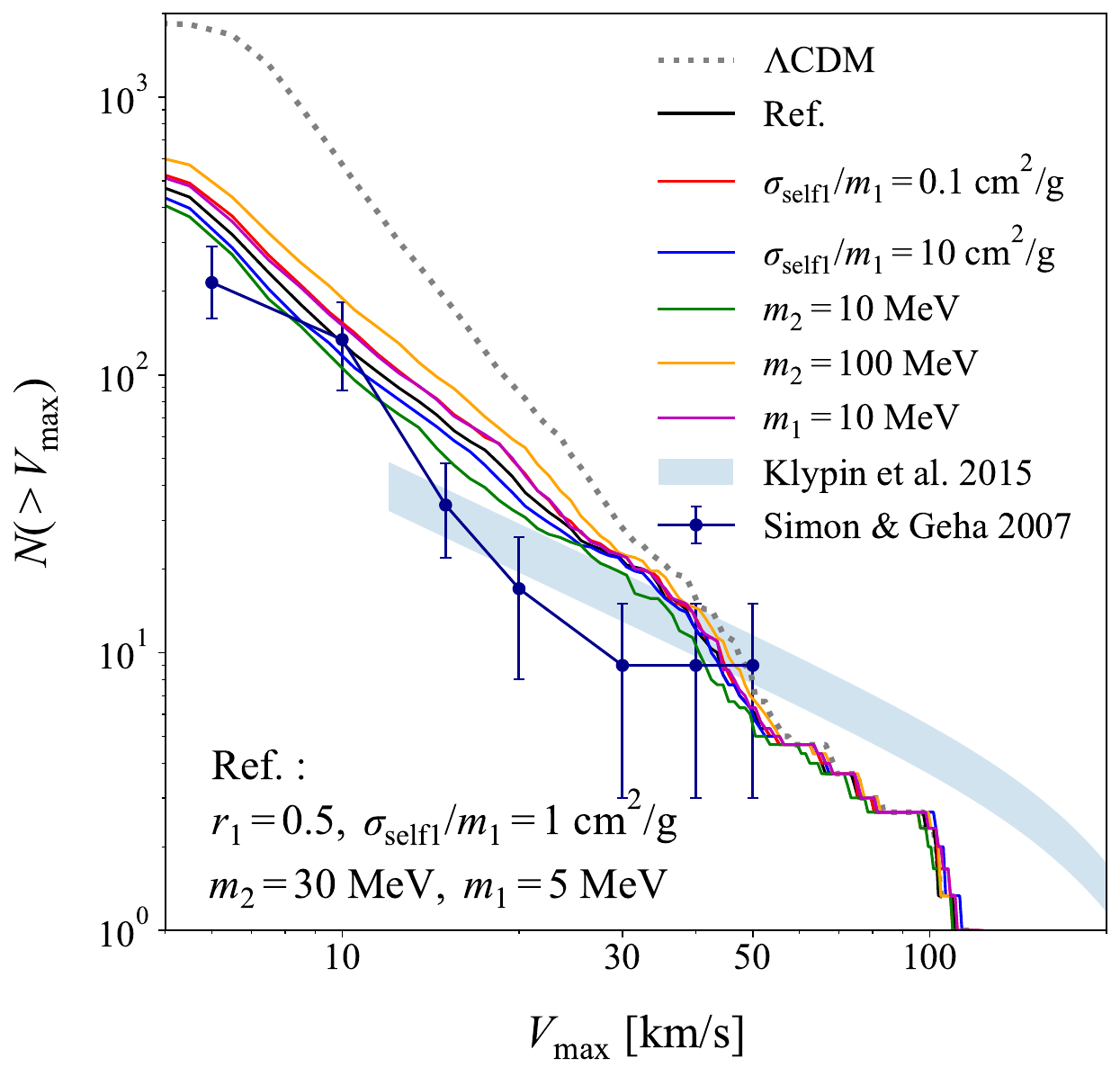}
    \caption{The number of sub-halos as a function of the maximum circular velocity, $V_{\rm max}$, for different values of $r_1$ (left) and $\sigma_{\text{self}1}/m_1$ or DM masses, with respect to the reference parameter set.     \label{fig:MCVF} }
\end{figure}

We explore sub-halo mass distributions via the 
maximum circular velocity function (MCVF) as shown in Figure \ref{fig:MCVF}. The MCVF displays a cumulative distribution of the number of sub-halos, $N(>V_{\rm max})$, based on the maximum circular velocity, $V_{\rm max}$, which is proportional to the corresponding sub-halo mass. The heavier sub-halos are associated with the larger value of $V_{\rm max}$. Blue points with error-bars show the local group observational data from Ref.~\cite{Simon_2007}, and the data in the blue (shaded) strip is taken from Ref.~\cite{2015MNRAS.454.1798K}.
The $\Lambda$CDM simulation shows a notable presence of small sub-halos within 
$V_{\rm max} \lesssim 30-40\rm{km/s}$.
Conversely, as $r_1$ increases in the two-component DM simulation (as shown in the left panel), the number of small sub-halos decreases. The most prominent reduction occurs at around $r_1 \sim 0.5$. However, further increasing $r_1$ results in a slight uptick in the number of small sub-halos.
Once again, this observation is in line with the trends shown by the linear power spectrum in Figure \ref{fig:Linear-PK} for $k$ modes of $\mathcal{O}(50~h{\rm Mpc^{-1}})$ that roughly correspond to the size of minihalos.

The right panel in Figure \ref{fig:MCVF} shows how the MCVF changes due to variations of the self-interaction $\sigma_{\text{self}1}/m_1$ and DM masses.
With an increase in $\sigma_{\text{self}1}/m_1$, the number of small sub-halos decreases since the larger value of $\sigma_{\text{self}1}/m_1$ maintains the self-heating effect for an extended period of time.
On the other hand, the heavier $\chi_2$ (lighter $\chi_1$) particles weaken (strengthen) the self-heating effect, which leads to an increase in the number of small sub-halos (see Eq. (\ref{eq:heat}).).

While the presence of baryonic effects might lower the magnitude of MCVF~\cite{Lovell:2016nkp,Kim:2021zzw} in the CDM only simulations, these effects may not be substantial to explain the observed data.
One way to reduce the discrepancy is to incorporate WDM species~\cite{Lovell:2016nkp,Kim:2021zzw}, and the light DM $\chi_1$ in our two-component model plays exactly the same role as WDM.

\subsection{Galactic Density Profile}
\label{sec:density_profile}

\begin{figure}[t]
    \centering
    \includegraphics[width=0.49\linewidth,clip]{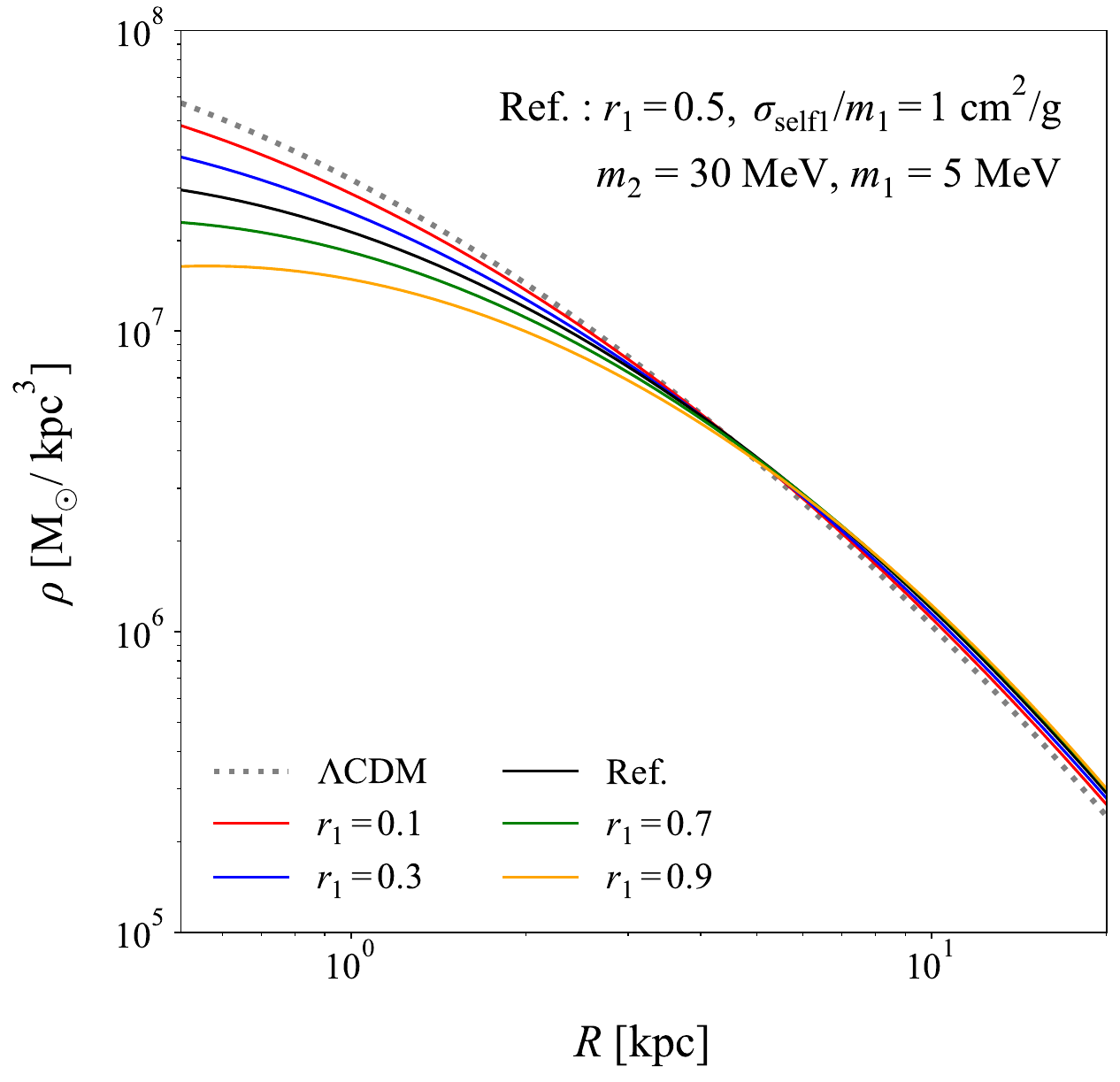}
    \includegraphics[width=0.49\linewidth,clip]{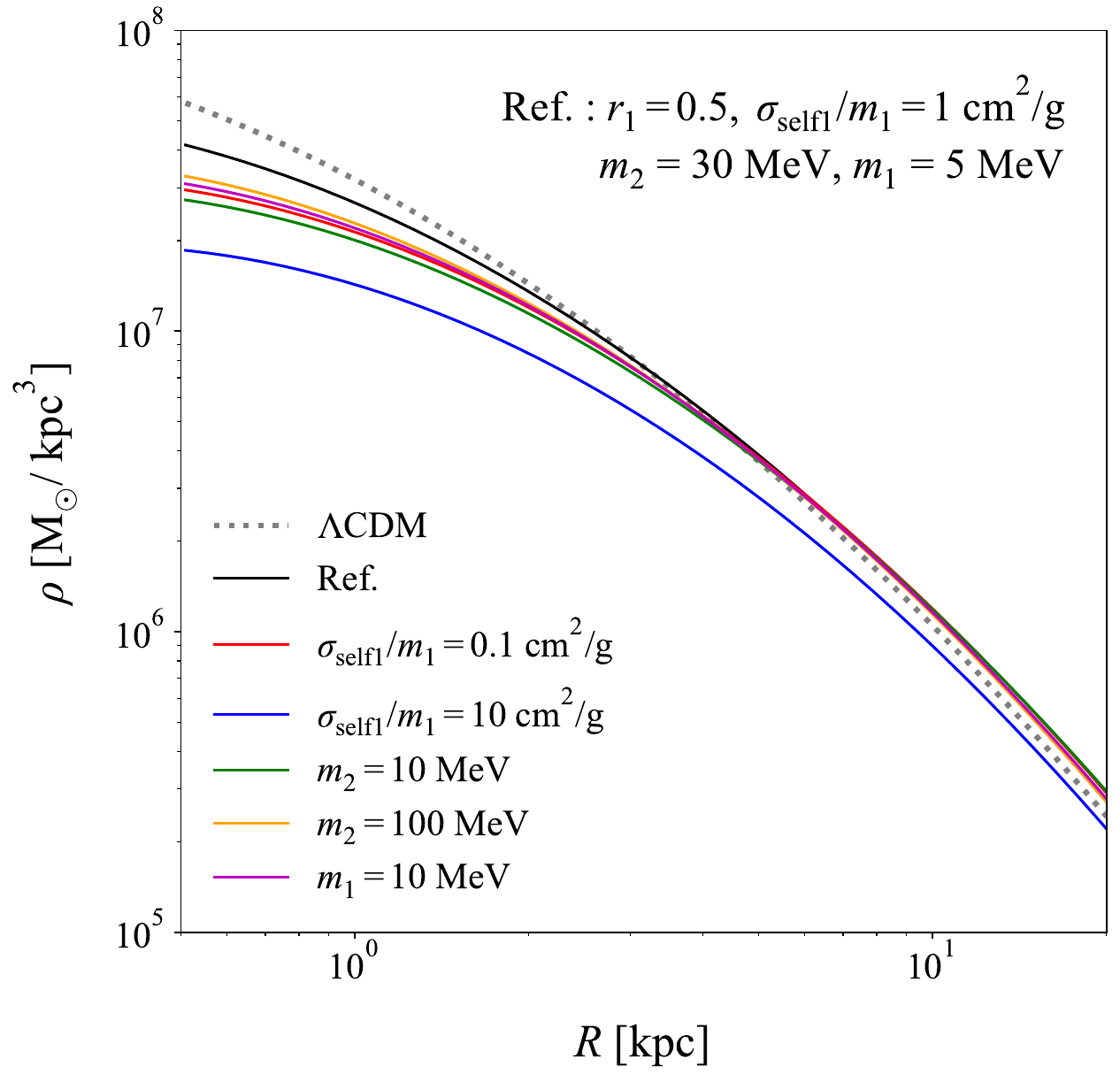}
    \caption{
    Total density profiles of halos with masses greater than $10^{10} M_{\odot}$ for different values of $r_1$ (left) and $\sigma_{\text{self}1}/m_1$ or DM masses (right) with respect to the reference parameter set. Results are averaged over four independent simulations. 
        }
    \label{fig:density_tot}
\end{figure}

\begin{figure*}[t]
\centering
\includegraphics[width=0.49\linewidth,clip]{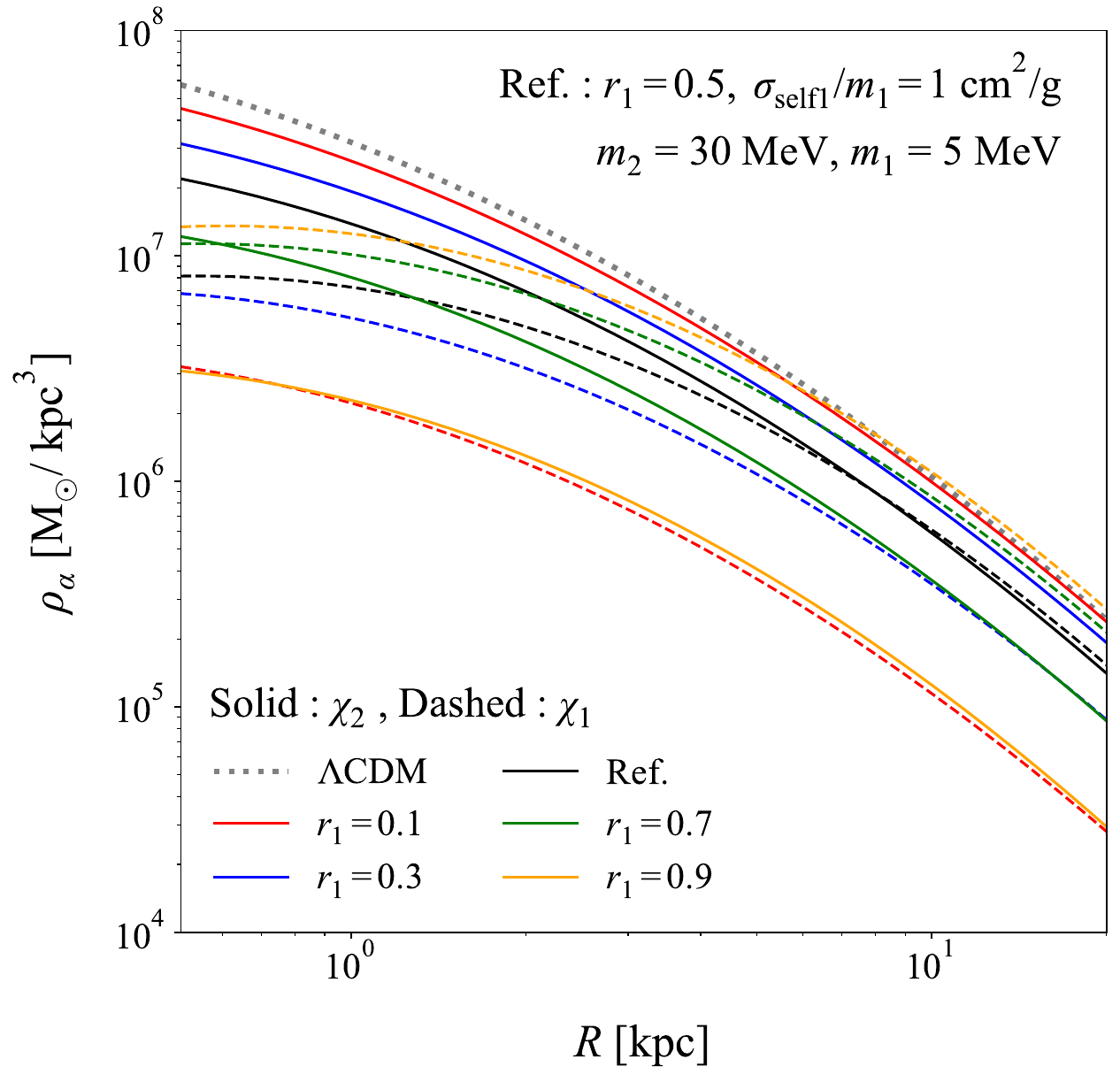}
\includegraphics[width=0.49\linewidth,clip]{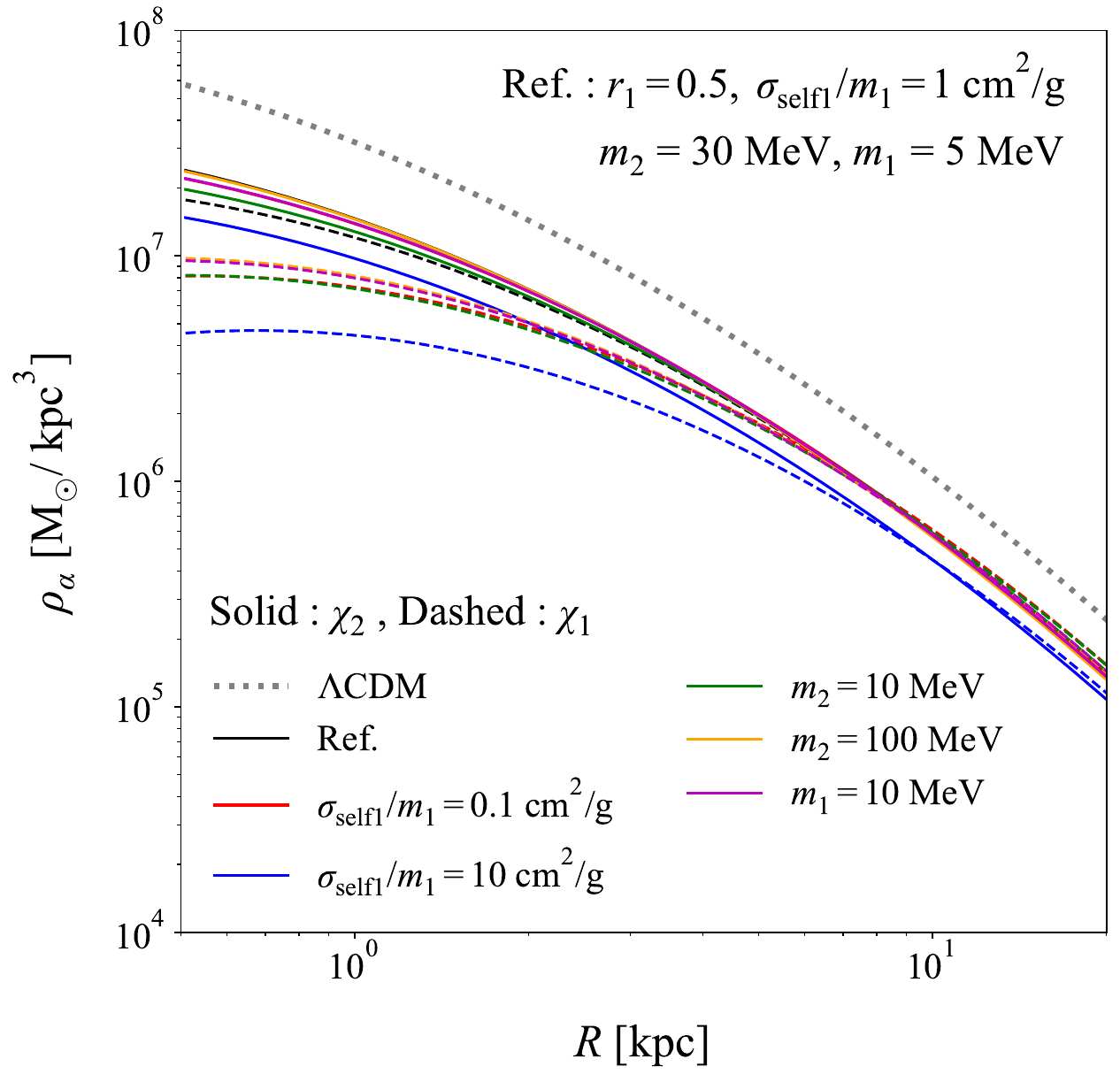} \\
\includegraphics[width=0.49\linewidth,clip]{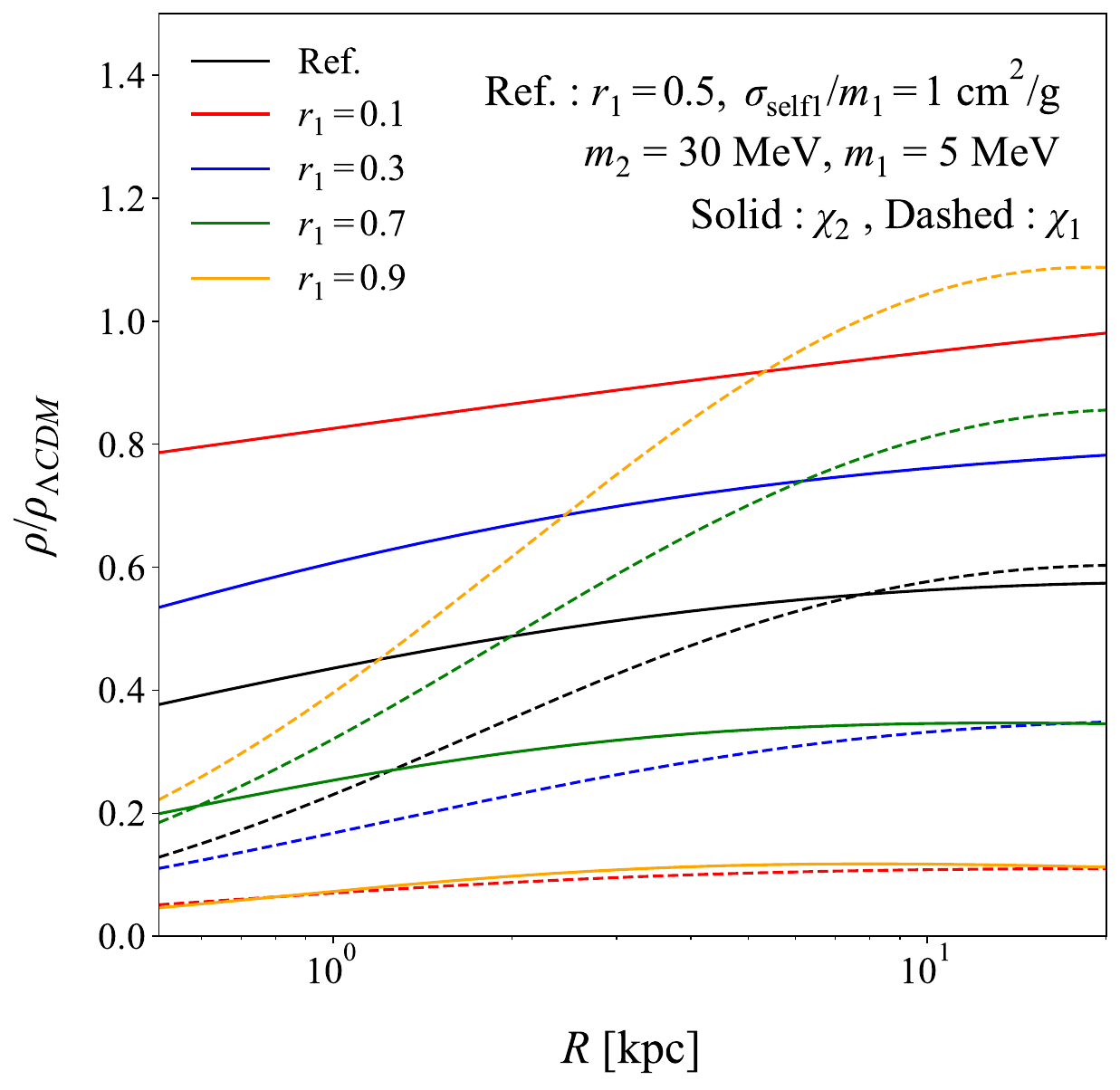}
\includegraphics[width=0.49\linewidth,clip]{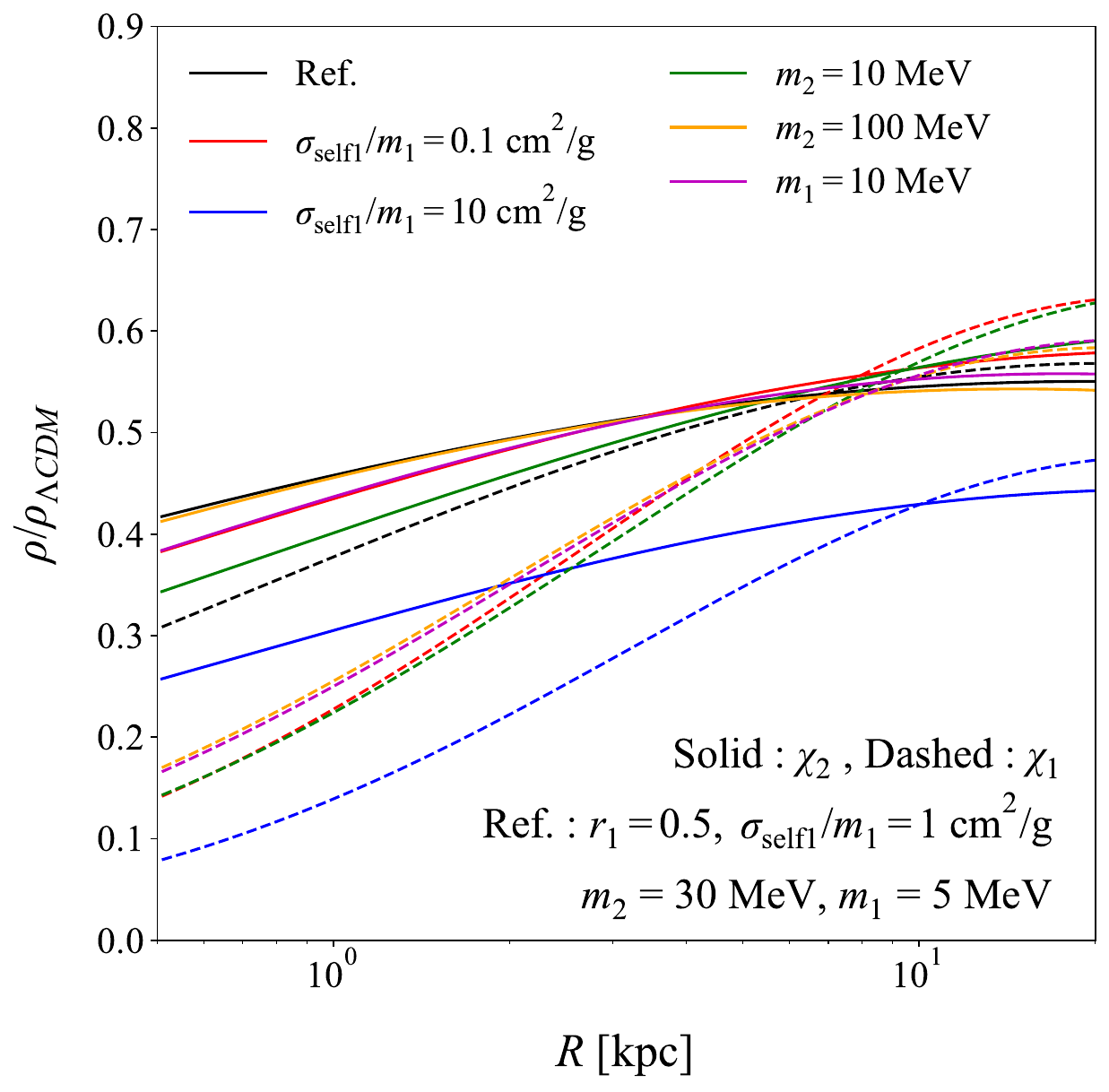}
\caption{Averaged individual $\chi_1$ (dashed) and $\chi_2$ (solid) density profiles of halos with masses greater than $10^{10} M_{\odot}$ for various parameters (top panels), and their ratios with respect to the $\Lambda$CDM profile (bottom panels).
}
\label{fig:density_sep} 
\end{figure*}

To further investigate the smaller substructures of halos, we take the average of the halo density profiles for masses greater than $10^{10} M_{\odot}$.
We run four independent $N$-body simulations with different random seeds for each parameter set, and then average the results to determine final density profiles.
Figure \ref{fig:density_tot} (left) shows the total density profiles for different values of $r_1$, while keeping all other parameters at their reference values.
The $\Lambda$CDM profile (gray dotted line) shows a similar power law behavior as
the Navarro-Frenk-White (NFW) density profile~\cite{Maccio:2008pcd}
\begin{equation}
    \rho_{\rm NFW}(r)=\frac{\rho_{0}}{(r/r_{sc}) \left(1+r/r_{sc}\right)^{2}} \;,
\label{eq:NFW}
\end{equation}
where the parameter $\rho_0$ and a scale radius $r_{sc}$ may vary among different halos~\cite{Correa:2015dva}. The NFW profile exhibits a power law of $\rho_{\rm NFW}(r) \sim r^{-3}$ at distances beyond $r_{sc}$, but it switches to $\rho_{\rm NFW}(r) \sim r^{-1}$ at distances smaller than $r_{sc}$.

On the other hand, the inner density profiles of two-component DM exhibit a more gradual slope and show reduced concentration (referred to as a core).
The core has a radius of $\sim 4 ~{\rm kpc}$, following a power law of
$\rho(r) \sim r^{-1/2}$. 
With increasing $r_1$ values, the core becomes shallower, as predicted by the linear power spectrum at $k \gg 100~h{\rm Mpc^{-1}}$ in Figure \ref{fig:Linear-PK}. The $\chi_1$ self-scattering implemented in the two-component DM simulations also plays a role in this effect, aiding in the formation of the cores as $r_1$ values increase. The core formation is consistent with observations in SIDM $N$-body simulations~\cite{Rocha:2012jg, Ahn:2004xt, Colin_2002, Moore:2000fp, Yoshida_2000, Spergel:1999mh}.

The right panel in \ref{fig:density_tot} shows how the density profiles change due to variations of the self-interaction $\sigma_{\text{self}1}/m_1$ and DM masses. With an increase in $\sigma_{\text{self}1}/m_1$, the density profiles decreases. 
On the other hand, the heavier $\chi_2$ (lighter $\chi_1$) particles weakens (strengthens) the self-heating effect, which lead to an increase in the density profiles.

Figure \ref{fig:density_sep} shows the individual $\chi_1$ and $\chi_2$ density profiles (top panel) and their ratios with respect to the $\Lambda$CDM profile (bottom panels) for various parameters. In the central region, $\chi_1$ densities are $\sim 70\%$ lower than the $\Lambda$CDM prediction, because the $\chi_1$ self-scattering blows away $\chi_1$ clumps outward. 
When $r_1 \sim 0.9$, the $\chi_1$ density in the outer region at $r \sim 10~{\rm kpc}$ can be $\sim 20\%$ higher than the $\Lambda$CDM prediction. At this length scale, the pressure due to the $\chi_1$ self-interaction balances out with gravity.
On the other hand, $\chi_2$ densities exhibit plateau distributions,
yet they gradually become shallower towards the central region 
because of the gravitational pull from the $\chi_1$ clumps.

\section{Summary and Outlook}
\label{sec:summary}

In this paper, we investigated cosmological aspect of a model with two-component CDM, where the lighter component behaves like WDM. For the first time, we explored the temperature evolution (section \ref{sec:temperature}), density perturbations (section \ref{sec:pertubation}), power spectrum (sections \ref{sec:linear} and \ref{sec:powerspectrum}), and maximum circular velocity function (section \ref{sec:subhalo}), and density profiles (section \ref{sec:density_profile}) using $N$-body simulations. These diverse cosmological implications provide crucial insights for particle physics experiments. 
such as Super-Kamiokande \cite{Super-Kamiokande:2002weg}, XENONnT \cite{XENON:2017lvq}, and JUNO \cite{JUNO:2021vlw}.

\begin{figure*}[t]
\centering
\includegraphics[width=0.99\linewidth,clip]{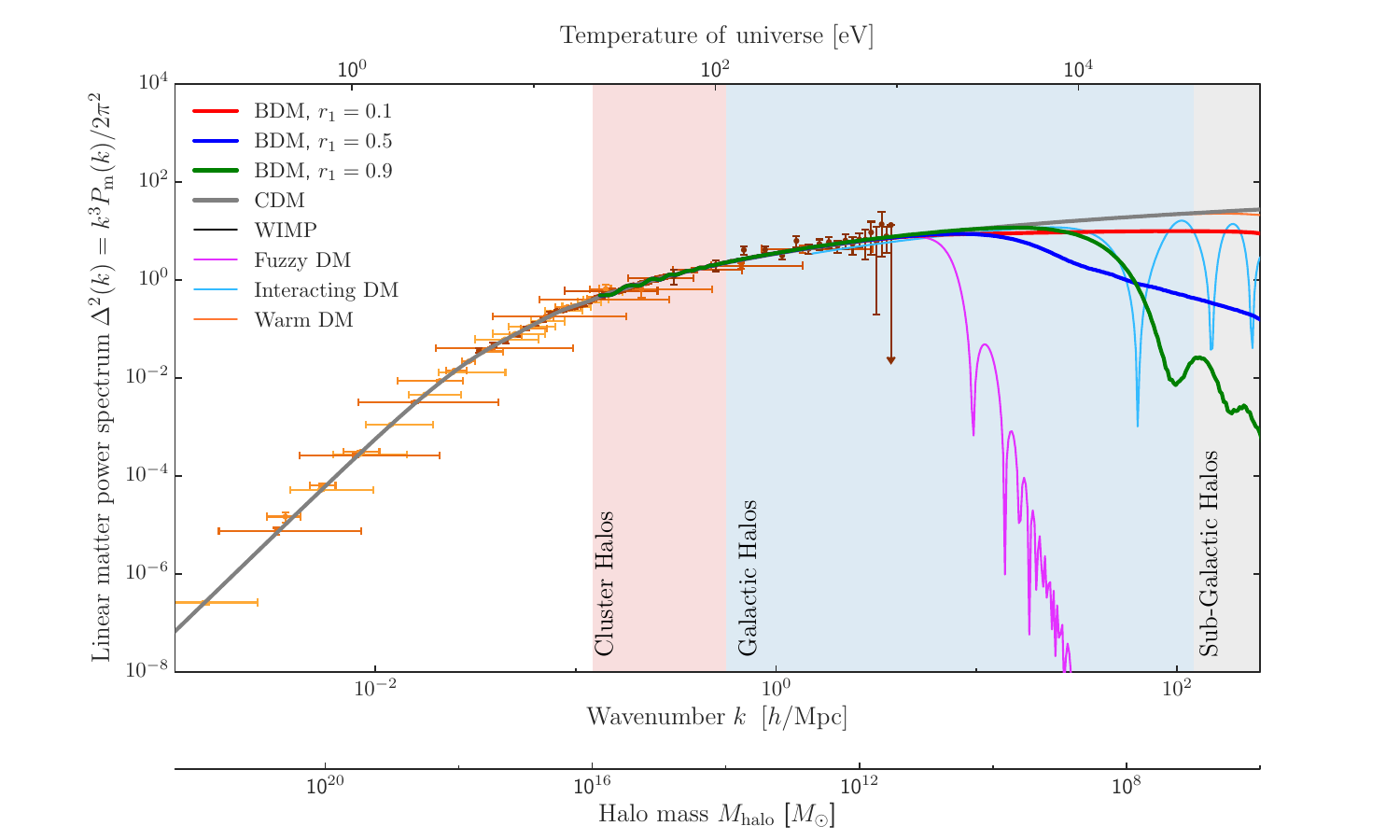}
\caption{
The dimensionless linear matter power spectrum extrapolated to $z = 0$. The BDM curves are represented by three solid curves (red for $r_1=0.1$, blue for $r_1=0.5$ and green for $r_1=0.9$).
All other curves are from Refs. \cite{Drlica-Wagner:2022lbd,Bechtol:2022koa}. 
The shaded vertical bands indicate the characteristic kinds of halos formed on each scale, and the horizontal axes indicate wavenumber, halo mass, and the temperature of the Universe when that mode entered the horizon.
}
\label{fig:Non-Pk} 
\end{figure*}

To compare the cosmology of the two-component BDM model against that of different DM models, we show the dimensionless linear matter power spectrum (see Figure \ref{fig:Non-Pk})
extrapolated to $z = 0$, following Refs. \cite{Drlica-Wagner:2022lbd,Bechtol:2022koa}. 
BDM results are shown as three solid curves (red for $r_1=0.1$, blue for $r_1=0.5$, and green for $r_1=0.9$).
All other curves are obtained from Refs. \cite{Drlica-Wagner:2022lbd,Bechtol:2022koa}. 
The shaded vertical bands roughly indicate the characteristic kinds of halos formed on each scale, and the horizontal axes indicate wavenumber, halo mass, and the temperature of the Universe when that mode entered the horizon.

Further investigation is required to fully explore the cosmological and phenomenological implications of BDM, including aspects such as baryonic effects, and gravitational wave signals. These topics are beyond the scope of the current study and will be addressed in future work \cite{KKLP}.

\appendix
\section{Perturbed Boltzmann Equations in Two-Component Boosted DM} 
\label{sec:appendixA}

In this section, we simplify the coupled perturbation equations in Eqs.~(\ref{eq:Perturbation1}--\ref{eq:Perturbation4}, \ref{eq:Einstein0s}--\ref{eq:Einsteinss}) by taking a special limit to make the physical meaning of the equations more easily understandable.
In the matter-dominated era, when $T \ll \mathcal{O}(\text{MeV})$, the background equilibrium densities, $\bar{\rho}_{1, \text{eq}}$ and $\bar{\rho}_{2, \text{eq}}$, can be neglected as they are suppressed by $\sim e^{-m_{1,2}/T}$. 
Their ratio, $\bar{\rho}_{2, \text{eq}}/\bar{\rho}_{1, \text{eq}}$, can also be neglected as we consider the mass difference $\delta m$ on the order of $\mathcal{O}(\text{MeV})$. 
During matter domination, the gravitational potentials are nearly constant on all scales, and hence we can neglect their time derivatives as well.
We will also assume that $\big<\sigma v \big>_{2 2\rightarrow 1 1}$ and $\big<\sigma v \big>_{1 1\rightarrow X X}$ are time-independent, and 
the process $\chi_1 \chi_ 1\rightarrow X X$ is a $p$-wave annihilation, since the s-wave annihilation is strongly constrained \cite{Kamada:2021muh}.

In addition, we will use the non-relativistic time-time component of Einstein equation $\nabla^2 \Phi  \simeq 4 \pi G a^2(\delta_1 \bar{\rho}_1 + \delta_2 \bar{\rho}_2 )$ and the space-space component of Einstein equation $\Phi \simeq \Psi$ with neglecting the anisotropic stress tensor.
Combining the time derivative of continuity Eqs.~(\ref{eq:Perturbation1}, \ref{eq:Perturbation3}) with both Euler Eqs.~(\ref{eq:Perturbation2}, \ref{eq:Perturbation4}) and the Einstein equations,
we obtain the system of coupled harmonic oscillators in Eqs.~(\ref{eq:2ndDelta1}, \ref{eq:2ndDelta2}).
When deriving this, time derivatives of background quantities are replaced by the zero-order Boltzmann equations in Eqs.~(\ref{eq:BE01}, \ref{eq:BE02}).

\begin{flalign}
    \frac{d^2 \delta_{2}}{dt^2} +\Big( 2 H + { \frac{\big<\sigma v \big>_{2 2\rightarrow 1 1}}{m_{2} } \bar{\rho}_{2} } \Big) { \frac{d \delta_{2}}{dt} } &- \Bigg\{ \frac{\big<\sigma v \big>_{2 2\rightarrow 1 1}}{m_{2} } H + \Big(\frac{\big<\sigma v \big>_{2 2\rightarrow 1 1}}{m_{2} } \Big)^2  \bar{\rho}_{2} + 4 \pi G \Bigg\} \bar{\rho}_{2} \delta_{2}  \nonumber \\
  & = 4 \pi G \bar{\rho}_{1} \delta_{1} + \frac{\big<\sigma v \big>_{2 2\rightarrow 1 1}}{m_{2} } \Big( H + \frac{\big<\sigma v \big>_{2 2\rightarrow 1 1}}{m_{2} } \bar{\rho}_2  \Big) \bar{\rho}_2  \Phi   \;,\label{eq:2ndDelta1} 
\end{flalign}
\begin{flalign}
    \frac{d^2 \delta_{1}}{dt^2} &+\Big( 2 H + {2\frac{\big<\sigma v \big>_{2 2\rightarrow 1 1}}{m_{2} } \frac{\bar{\rho}^2_{2}}{\bar{\rho}_{1}} } + \frac{\big<\sigma v \big>_{1 1\rightarrow XX}}{m_{1} } \bar{\rho}_{1} \Big) {\frac{d \delta_{1}}{dt} }   \nonumber \\
     &+ \Bigg\{ -\frac{\big<\sigma v \big>_{2 2\rightarrow 1 1}}{m_{2} } \frac{\bar{\rho}^2_{2}}{\bar{\rho}_{1}}  \Big(H + \frac{\big<\sigma v \big>_{2 2\rightarrow 1 1}}{m_{2} } \frac{\bar{\rho}^2_{2}}{\bar{\rho}_{1}}  \Big( 2 \frac{\bar{\rho}_1}{\bar{\rho}_2}   + \frac{m_1}{m_2}  -1 \Big)  \Big)  \label{eq:2ndDelta2}  \\
   & \hspace*{0.8cm} - \frac{\big<\sigma v \big>_{1 1\rightarrow XX}}{m_{1} }  \bar{\rho}_{1} \Big( H + \frac{\big<\sigma v \big>_{1 1\rightarrow XX}}{m_{1} }  \bar{\rho}_{1}  \Big) - 4 \pi G  \bar{\rho}_{1}   \nonumber \\
    & \hspace*{0.8cm}+ \frac{\big<\sigma v \big>_{2 2\rightarrow 1 1}}{m_{2} } \frac{\big<\sigma v \big>_{1 1\rightarrow XX}}{m_{1} }  \bar{\rho}^2_{2} \Big( 2 + \frac{m_1}{m_2} \Big)  + c^2_{s,1} \frac{k^2}{a^2}  \Bigg\} \delta_{1}   \nonumber \\
   & =  { 3\frac{\big<\sigma v \big>_{2 2\rightarrow 1 1}}{m_{2} } \frac{\bar{\rho}^2_{2}}{\bar{\rho}_{1}} \frac{d \delta_{2}}{dt} }
   -  \Bigg\{ 2 H \frac{\big<\sigma v \big>_{2 2\rightarrow 1 1}}{m_{2} } \frac{\bar{\rho}^2_{2}}{\bar{\rho}_{1}}   + \Big(\frac{\big<\sigma v \big>_{2 2\rightarrow 1 1}}{m_{2} } \Big)^2 \Big( 3 \frac{\bar{\rho}^3_{2}}{\bar{\rho}_{1}} + 2 \Big( \frac{m_1}{m_2} - 1 \Big) \frac{\bar{\rho}^4_{2}}{\bar{\rho}^2_{1}} \Big)  \nonumber \\
   & \hspace*{4.5cm}- 2 \frac{\big<\sigma v \big>_{2 2\rightarrow 1 1}}{m_{2} } \frac{\big<\sigma v \big>_{1 1\rightarrow XX}}{m_{1} } \bar{\rho}^2_{2} - 4 \pi G \bar{\rho}_2 \Bigg\}   \delta_{2} \nonumber \\  
   &+ \Bigg\{ - \frac{\big<\sigma v \big>_{2 2\rightarrow 1 1}}{m_{2} } \Big( H \frac{\bar{\rho}^2_{2}}{\bar{\rho}_{1}}  + \frac{\big<\sigma v \big>_{2 2\rightarrow 1 1}}{m_{2} } \Big(  \frac{\bar{\rho}^3_{2}}{\bar{\rho}_{1}} + 2 \Big( \frac{m_1}{m_2} - 1 \Big) \frac{\bar{\rho}^4_{2}}{\bar{\rho}^2_{1}} \Big) \Big) \nonumber \\
   &\hspace*{0.8cm}
   + \frac{\big<\sigma v \big>_{1 1\rightarrow XX}}{m_{1} } \Big( H \bar{\rho}_{1} + \frac{\big<\sigma v \big>_{1 1\rightarrow XX}}{m_{1} } \bar{\rho}^2_{1} \Big) 
- \frac{\big<\sigma v \big>_{2 2\rightarrow 1 1}}{m_{2} } \frac{\big<\sigma v \big>_{1 1\rightarrow XX}}{m_{1} } \frac{m_1}{m_2} \bar{\rho}^2_2  \Bigg\}  \Phi  \;, \nonumber 
\end{flalign}

\begin{flalign}
   k^2_J \simeq \frac{a^2}{c^2_{s,1}} \Bigg( &\frac{\big<\sigma v \big>_{2 2\rightarrow 1 1}}{m_{2} } \frac{\bar{\rho}^2_{2}}{\bar{\rho}_{1}}  \Big(H + \frac{\big<\sigma v \big>_{2 2\rightarrow 1 1}}{m_{2} } \frac{\bar{\rho}^2_{2}}{\bar{\rho}_{1}}  \Big( 2 \frac{\bar{\rho}_1}{\bar{\rho}_2} 
   + \frac{m_1}{m_2}  -1 \Big)  \Big) \nonumber \\ 
   &+ \frac{\big<\sigma v \big>_{1 1\rightarrow XX}}{m_{1} }  \bar{\rho}_{1} \Big( H + \frac{\big<\sigma v \big>_{1 1\rightarrow XX}}{m_{1} }  \bar{\rho}_{1}  \Big) \label{eq:kJ}  \\ 
   &
   - \frac{\big<\sigma v \big>_{2 2\rightarrow 1 1}}{m_{2} } \frac{\big<\sigma v \big>_{1 1\rightarrow XX}}{m_{1} }  \bar{\rho}^2_{2} \Big( 2 + \frac{m_1}{m_2} \Big) + 4 \pi G  \bar{\rho}_{1} \Bigg)   \;.  \nonumber
\end{flalign}

In the homogeneous part of Eq.~(\ref{eq:2ndDelta1}), the density contrast $\delta_2$ grows due to the source of gravity because the coefficient of the $\delta_2$ term is positive.
However, it experiences the additional friction, the third term on the left-hand side of the equation, owing to the disappearance of the gravitational potential well by the $\chi_2$ annihilation. 
This implies that in the large $r_1$ limit where $\big<\sigma v \big>_{2 2\rightarrow 1 1}$ is sizable, the friction increases, leading to a decrease in the growth of $\delta_2$. 
This attribute of $\chi_2$ sets it apart from conventional CDM. 
The presence of inhomogeneous terms on the right-hand side signifies the interconnected nature of $\delta_2$ with $\delta_1$ and gravity $\Phi$.

In contrast, when examining only the homogeneous part of Eq.~(\ref{eq:2ndDelta2}), $\chi_1$ density contrast may oscillate or increase depending on the magnitude of the pressure term, $c^2_{s,1} k^2/a^2$. 
This is because the competition between gravity and the pressure term dictates the overall sign of the $\delta_1$ term.
If the sign is positive (negative), then the $\chi_1$ perturbation oscillates (grows).
According to Eq.~(\ref{eq:cs}), the pressure increases with the temperature $T_1$. 
For example, a small $r_1$ value leads to a large $T_1$. 
If the $T_1$ is large enough, it can generate a pressure that dominates gravity and induces oscillatory behavior in the $\chi_1$ perturbation.
We can define a wavenumber $k$, called the Jeans scale in Eq.~(\ref{eq:kJ}), at which the oscillation frequencies vanish. 
For a reference set of parameters ($r_1=0.5$, $\sigma_{\text{self}1}/m_1 = 1~\text{cm}^2/\text{g}$, $m_2 = 30$ MeV, $m_1 = 5$ MeV), the corresponding Jeans scale, at the time of matter-radiation equality, is given by $k_J \simeq 30 ~h{\rm/Mpc^{-1}}$, which predicts the onset of oscillations at very small scales.

\acknowledgments
We thank Ethan Nadler for sharing the dimensionless linear matter power spectrum data for different dark matter models.
We thank Donghui Jeong, Jinn-Ouk Gong and Chang Sub Shin for useful discussion. 
The work is supported by the National Research Foundation of Korea grant funded by the Korea government (MSIT) [NRF-2021R1C1C1005076 (JHK, SHL),  RS-2024-00356960 (JCP)].
KK is supported in part by US DOE DE-SC0024407 and University of Kansas General Research Fund allocation.

\bibliographystyle{JHEP}
\bibliography{draft}

\end{document}